\documentclass[10pt,a4paper,twocolumn,english,prb,aps,showpacs,floatfix,groupedaddress,superscriptaddress]{revtex4-1}

\usepackage{graphicx}
\usepackage{epsfig}
\usepackage[english]{babel}
\usepackage{amsmath}
\usepackage{amssymb}
\usepackage{amsfonts}
\usepackage{longtable}
\usepackage{dcolumn} 
\usepackage{bm}
\usepackage{bbm}
\usepackage{nicefrac}
\usepackage{color,array}
\usepackage{colortbl}
\usepackage{verbatim}
\newcommand{\kvec}{\textbf{k}}	
\newcommand{\qvec}{\textbf{q}}

\newcommand{\crea}[2]{#1^\dagger_{#2}}			 

\newcommand{\EW}[1]{\left\langle#1\right\rangle}
\newcommand{\abl}{\frac{\text{d}}{\text{d}t}}

\begin{document}
\title{Signatures of nonadiabatic BCS state dynamics in pump-probe conductivity}

\author{H. Krull}
\email{holger.krull@tu-dortmund.de}
\affiliation{Lehrstuhl f\"{u}r Theoretische Physik I, Technische Univerit\"at Dortmund, 
Otto-Hahn Stra\ss{}e 4, 44221 Dortmund, Germany}
\author{D. Manske}
\email{d.manske@fkf.mpg.de}
\affiliation{Max-Planck-Institut f\"{u}r Festk\"{o}rperforschung,
Heisenbergstra\ss{}e 1, D-70569 Stuttgart, Germany}
\author{G. S. Uhrig}
\email{goetz.uhrig@tu-dortmund.de}
\affiliation{Lehrstuhl f\"{u}r Theoretische Physik I, Technische Univerit\"at Dortmund, 
Otto-Hahn Stra\ss{}e 4, 44221 Dortmund, Germany}
\author{A. P. Schnyder}
\email{a.schnyder@fkf.mpg.de}
\affiliation{Max-Planck-Institut f\"{u}r Festk\"{o}rperforschung,
Heisenbergstra\ss{}e 1, D-70569 Stuttgart, Germany}

\date{\rm\today}

\begin{abstract}
We theoretically study the pump-probe response of nonequilibrium BCS superconductors coupled to optical phonons. 
For ultrashort pump pulses a nonadiabatic regime emerges, which is characterized by oscillations of the superconducting order parameter
as well as by the generation of coherent phonons. Using the density-matrix formalism, we compute the pump-probe response in the nonadiabatic regime of the coupled Bogoliubov quasiparticle-phonon system and determine the signatures of the order parameter and the phonon oscillations in the pump-probe conductivity. We find that  the nonadiabatic dynamics of the BCS superconductor 
reflects itself in oscillations of the pump-probe response as  functions of delay time  $\delta t$
between  pump and probe pulses. We argue that from the analysis of this oscillatory behavior both   frequency and   decay time of the algebraically decaying order-parameter oscillations can be inferred. 
Similarly, the coherent phonons are evidenced in the pump-probe conductivity by oscillations with the frequency of the phonons. 
Remarkably, we find that the oscillatory response in the pump-probe conductivity is resonantly enhanced when the frequency of the order-parameter oscillations is tuned to the phonon energy. 
\end{abstract}


\pacs{74.40.Gh, 63.20.kd, 78.47.J-, 78.20.Bh}


\maketitle

\section{Introduction}

The nonequilibrium response of superconductors 
has been the subject of considerable interest for a long  time~\cite{kopninBook,Orenstein12}.
In the past few years, fueled by recent advances in femtosecond terahertz (THz) laser technology~\cite{Fulop12},
many experimental~\cite{matsunagaPRL13,matsunagaPRL12,mansartPNAS13,Beck2013,DalScience2013,beckDemsarPRL11,Rett10,leitensdorferPRL10,Mans09,rubhausenPRL09,PerfettiPRL07,Kain05,Ave01,kaindlScience00} 
and theoretical~\cite{volkov1974,amin2004,barankov2004,Yuz05,Yuz06,nicolePRB03,kabanovPRL05,Unter08,Papen07,Papen08,Papen09,Schny11,Akba13,zachmannNJP13} studies have focused on the ultrafast time response of the BCS state subjected to nonadiabatic perturbations. For example, Matsunaga \textit{et al.}\cite{matsunagaPRL12,matsunagaPRL13} have employed THz pump-THz probe spectroscopy to investigate the nonadiabatic dynamics of  superconducting thin films after the  injection of Bogoliubov quasiparticles with energies just above the gap edge. 
 Transient oscillations with the frequency of the amplitude mode of the superconducting order parameter have been observed in these measurements.
Besides pump-probe experiments, also tunneling spectroscopy has recently been used to study 
the ultrafast response of superconducting tunnel junctions~\cite{ponomarev2013}. 

On the theory side, intensive efforts have been devoted to  the analysis of  persistent and damped oscillations in the Bogoliubov quasiparticle occupations, 
which are induced by either an interaction quench~\cite{volkov1974,amin2004,barankov2004,Yuz05,Yuz06} or by ultrafast photoexcitations~\cite{Akba13,zachmannNJP13,Schny11,Papen07,Papen08,Papen09}. 
In particular, it has been shown that  excitation pulses with pulse duration $\tau_{\textrm{p}}$, see Eq.~\eqref{pulseShape}, shorter than the inverse BCS gap energy $| \Delta |$, i.e., $\tau_{\textrm{p}} \ll h / ( 2 | \Delta | )$, generate coherent oscillations in the order parameter amplitude and in the quasiparticle occupations with frequency  $\omega_{\Delta_{\infty}}  \approx  2 |\Delta | / \hbar $~\cite{volkov1974,amin2004,barankov2004,Yuz05,Yuz06}.  Moreover, in the presence of an optical-phonon branch with energy of the order
of $\hbar \omega_{\textrm{ph}}  \approx  2 | \Delta | $, coherent  quasipersistent phonon oscillations can be generated~\cite{Schny11}. While
the creation of these coherent oscillations is well understood theoretically, it has remained an open question whether and how
this nonadiabatic BCS state dynamics could be observed in experimental pump-probe spectra~\cite{Papen08}.

In this paper, we numerically simulate the pump-probe response of  nonequilibrium BCS superconductors coupled to optical-phonon modes and determine the signatures of the coherent Cooper pair and phonon dynamics in the pump-probe conductivity. 
 Our aim is to provide systematic calculations of the pump-probe response, such that experimental measurements can 
 more easily be interpreted and compared to the theoretical predictions.
Using density-matrix theory~\cite{Ross02}, we derive equations of motion for the coupled Bogoliubov quasiparticle-phonon system treating the phonons at a  fully quantum kinetic level. 
The pump-pulse-induced dynamics of this model is investigated in the collisionless regime, i.e., at time scales shorter than the
quasiparticle relaxation time, for which BCS mean-field theory is applicable~\cite{volkov1974,YuzJofPhys05}.
 The pump-probe experiments of Matsunaga \textit{et al.}~\cite{matsunagaPRL12,matsunagaPRL13}  have shown
that in Nb$_{1-x}$Ti$_x$N thin films this collisionless regime lasts up to $10$~ps, considerably longer than the period of the order parameter oscillations, 
which is of the order of $1$ ps.

We study different hierarchies of the involved time scales, focusing for the most part on the case where the
pump-pulse length $\tau_{\textrm{p}}$ is much shorter than both the phonon period  $\tau_{\textrm{ph}} = ( 2 \pi ) / \omega_{\textrm{ph}} $  and
the dynamical time scale of the superconductor $\tau_{\Delta}  \approx h / ( 2 | \Delta | ) $. In this  regime, both the phononic and quasiparticle subsystems evolve in a nonadiabatic fashion, leading to order-parameter oscillations and  the creation of coherent phonons. We present a detailed analysis of the generic features in the pump-probe response resulting from
these oscillatory behaviors. In particular, it is shown that the coherent Cooper pair dynamics as well as 
the finite density of coherent phonons produce oscillations 
in the pump-probe conductivity as a function of the delay time  $\delta t$
between  pump and probe pulses (see Figs.~\ref{FigM1}-\ref{FigM4}). 
Interestingly, the pump-probe signal is resonantly enhanced and exhibits strong nondecaying oscillations 
when the frequency of 
the order-parameter oscillation is tuned to the phonon energy. This condition can be achieved by adjusting the integrated pump laser intensity  (see Fig.~\ref{FigM5}).

The remainder of this paper is organized as follows. In Sect.~\ref{Sec:ModHam} we introduce the microscopic Hamiltonian of the model and specify the parameters for our numerical simulations. The equations of motion are derived in Sect.~\ref{sec:method} 
using  density-matrix theory. Our numerical results are presented and analyzed in Sects.~\ref{sec:result} and~\ref{sec:Results_zwei}. Finally, we summarize and conclude our findings in Sect.~\ref{sec:conclusion}. 
Some technical details and additional plots are presented in Appendixes~\ref{sec:eom} and~\ref{sec:example_im}, respectively.

\section{Microscopic Hamiltonian} \label{Sec:ModHam}

We consider a single-band $s$-wave  superconductor coupled to an optical-phonon mode described by the Hamiltonian 
$H=H_{\text{sc}}+H_{\text{ph}}+H_{\text{el-ph}}$, where $H_{\text{sc}}$ represents the mean-field BCS Hamiltonian
\begin{subequations} \label{modHam}
\begin{align} \label{SChamm}
H_{\text{sc}}=\sum_{\textbf{k},\sigma}\epsilon_{\textbf{k}}c^{\dagger}_{\textbf{k},\sigma}c^{\phantom{\dagger}}_{\textbf{k},\sigma}-\sum_{\textbf{k}\in W}\left(\Delta c^{\dagger}_{\textbf{k},\uparrow}c^{\dagger}_{-\textbf{k}\downarrow}+\Delta^{*} c^{\phantom{\dagger}}_{-\textbf{k},\downarrow}c^{\phantom{\dagger}}_{\textbf{k},\uparrow}\right),
\end{align}
the free-phonon part $H_{\text{ph}}$ is given by 
\begin{align} \label{freePhon} 
H_{\text{ph}}=\sum_{\textbf{p}}\hbar\omega_{\text{ph}}\left(b^{\dagger}_{\textbf{p}}b^{\phantom{\dagger}}_{\textbf{p}}+\frac{1}{2}\right),
\end{align}
and $H_{\text{el-ph}}$ denotes the  interaction between  electrons and phonons 
\begin{align} \label{froHam}
H_{\text{el-ph}}= \frac{1}{\sqrt{N}} \sum_{\textbf{p},\textbf{k},\sigma}g_{\text{ph}}\left(b^{\dagger}_{-\textbf{p}}+b^{\phantom{\dagger}}_{\textbf{p}}\right)c^{\dagger}_{\textbf{k}+\textbf{p},\sigma}c^{\phantom{\dagger}}_{\textbf{k},\sigma},
\end{align}
\end{subequations}
where $N$ is the number of lattice sites. 
In Eq.~\eqref{SChamm} $c^{\phantom{\dagger}}_{\textbf{k},\sigma}$ ($c^{\dagger}_{\textbf{k},\sigma}$) represents the electron annihilation (creation) operator with momentum $\textbf{k}$ and spin $\sigma$, $\epsilon_{\textbf{k}}=\frac{\hbar^2\textbf{k}^2}{2m}-E_{\text{F}}$ is the electron dispersion relation, $m$ denotes the effective electron mass, and $E_{\text{F}}$ stands for the Fermi energy. The second sum in Eq.~\eqref{SChamm} is taken over the set $W$ of all $\textbf{k}$ vectors with $|\epsilon_{\textbf{k}}|\leq\hbar\omega_{\textrm{c}}$, $\omega_{\textrm{c}}$ being the frequency cutoff.  The superconducting order parameter $\Delta (t) $ is assumed to be of $s$-wave symmetry, with 
$
\Delta (t) =\frac{W_0}{N}\sum_{\textbf{k}\in W} \langle c^{\phantom{\dagger}}_{-\textbf{k},\downarrow}c^{\phantom{\dagger}}_{\textbf{k},\uparrow} \rangle,
$ and $W_0$ an attractive interaction constant. 

 We emphasize that we are generally looking for signatures of the dynamics after
the pump pulses  and, in particular, in the dynamics of certain low-lying phonons. The phonon mode~\eqref{freePhon} considered here does not generate the superconductivity. The pairing interaction is assumed to be mediated by other bosons at higher energies, i.e., for instance by spin fluctuations or by  phonons at high energies of the order of the Debye energy $\hbar \omega_{\mathrm{D}}\approx 30$meV.
This energy scale is much larger than the energy scale of
the Bogoliubov quasiparticles so that these other bosons influence the low-energy dynamics
only indirectly  via virtual processes, and hence, they do not need to be treated explicitly.
The corresponding  electron-boson couplings are assumed to be integrated out and enter in the mean-field treatment of superconductivity via $W_0$.

 For our numerical calculations we have to fix the parameters.
Motivated by the numbers for Pb \cite{pooleBook07} we fix the parameters as follows: gap in the initial state  $\Delta (t_{\textrm{i}} ) =1.35$ meV,  Fermi energy $E_{\textrm{F}}=9479$~meV,  energy cutoff $\hbar\omega_{\textrm{c}}=8.3$~meV, and effective electron mass $m=1.9m_0$, with $m_0$ being the free electron mass.
The operator $b^{\dagger}_{\textbf{p}}$ ($b^{\phantom{\dagger}}_{\textbf{p}}$) in Eq.~\eqref{freePhon} creates (annihilates) phonons with wave vector $\textbf{p}$ and frequency $\omega_{\text{ph}}$, where   $\omega_{\text{ph}}$ is assumed to be constant for simplicity.  Similarly, the coupling between  electrons and  phonons is taken to be of Holstein form [Eq.~\eqref{froHam}] 
with a momentum-independent interaction constant $g_{\text{ph}}$.

In the following, we study the nonequilibrium response of Hamiltonian~\eqref{modHam} to a short intense pump pulse which injects  a nonthermal 
distribution of Bogoliubov quasiparticles into the system. The considered pump pulse is of Gaussian shape with photon frequency $\omega_{\textrm{p}}$, photon wave vector ${\bf q}_{\textrm{p}}=|\textbf{q}_{\textrm{p}} | \hat{\textbf{e}}_x$, full width at half maximum (FWHM) $\tau_{\textrm{p}}$, and amplitude ${\bf A}_{\textrm{p}} =|\textbf{A}_{\textrm{p}} | \hat{\textbf{e}}_{y}$. Working within the Coulomb gauge, the pump pulse is expressed in terms of the transverse vector potential~\cite{footnote0},
\begin{align} \label{pulseShape}
\textbf{A}_{\textbf{q}}(t)=\textbf{A}_{\textrm{p}}  e^{-\left(\frac{2\sqrt{\ln{2}}t}{\tau_{\textrm{p}}}\right)^2}\left(\delta_{\textbf{q},\textbf{q}_{\textrm{p}} }e^{-i\omega_{\textrm{p}}  t}+\delta_{\textbf{q},-\textbf{q}_{\textrm{p}} }e^{i\omega_{\textrm{p}}  t}\right) .
\end{align}
Thereby, the coupling of the pump pulse to the superconductor is described by $H_{\text{em}}=H_{\text{em}}^{(1)}+H_{\text{em}}^{(2)}$, where
\begin{subequations}
\begin{align}
H_{\text{em}}^{(1)}&=\frac{e\hbar}{2m}\sum_{\textbf{k},\textbf{q},\sigma}(2\textbf{k}+\textbf{q})\textbf{A}_{\textbf{q}}(t)c^{\dagger}_{\textbf{k}+\textbf{q},\sigma}c^{\phantom{\dagger}}_{\textbf{k},\sigma},\\
H_{\text{em}}^{(2)}&=\frac{e^2}{2m}\sum_{\textbf{k},\textbf{q},\sigma}\left(\sum_{\textbf{q}^{\prime}}\textbf{A}_{\textbf{q}-\textbf{q}^{\prime}}(t)\textbf{A}_{\textbf{q}^{\prime}}(t)\right)c^{\dagger}_{\textbf{k}+\textbf{q},\sigma}c^{\phantom{\dagger}}_{\textbf{k},\sigma} .
\end{align}
\end{subequations}
We stress that the wave vector ${\bf q}_{\textrm{p}}$ must be kept finite in order to desribe
the effect of the pump pulse correctly. If it were set to zero, all linear couplings of the
electromagnetic field to the fermions are gone. This can be most clearly seen if we
consider the metallic case without superconductivity. The single-band model does not allow
any direct excitation process $\Delta {\bf q}=0$ but only indirect ones. Thus, neglecting
the finiteness of the wave vector prevents any excitation in the linear regime.
Moreover, certain effects such as the lowering of the order parameter $\Delta(t)$ and the Pauli blocking would be ill described
if we set ${\bf q}_{\textrm{p}}=0$~\cite{Papen07}.

The absorption spectrum of the nonequilibrium state is measured by a probe pulse, which follows the pump pulse after a certain delay time $\delta t$. The probe pulse has the same shape as the pump pulse, Eq.~\eqref{pulseShape}, but much weaker intensity. We consider both negative and positive pump-probe delay times $\delta t$, depending on whether the probe pulse precedes the pump pulse ($\delta t<0$) or follows after it ($\delta t>0$). 
For the numerical computations we assume that the pump pulse is centered in time at $t=0$ ps and has photon energy  $\hbar\omega_{\text{p}}=3$ meV, which is slightly larger than twice the  gap energy $2 | \Delta (t_{\textrm{i}} )  |$ of the superconductor in the initial state at $t=t_{\textrm{i}}$. The probe pulse is taken to be very short in time with FWHM  $\tau_{\text{pr}}=0.25$~ps and center energy $\hbar\omega_{\text{pr}}=2.5$ meV (see Fig.~\ref{FigM9_new} in Appendix~\ref{sec:eom}).

Thus, the probe pulse contains a broad range of frequencies which cover almost the
entire energy range of excited quasiparticles induced by the pump pulse similar
to recent experiments~\cite{DalScience2013,matsunagaPRL12,matsunagaPRL13}.
Nonlinear couplings between the superconductor and the probe pulse are neglected  
because probing is done with much weaker intensity in the linear response regime.

\section{Equations of motion}
\label{sec:method}

In order to simulate the pump-probe conductivity, we need to determine the temporal evolution of the electric current density  ${\bf j}_{{\bf q}_{\textrm{pr}}} (\delta t, t)$, where ${\bf q}_{\textrm{pr}}= \left| {\bf q}_{\textrm{pr}} \right| \hat{{\bf e}}_x$ is the 
wave vector of the probe pulse and
\begin{eqnarray} \label{defCurrent}
{\bf j}_{\textbf{q}_{\text{pr}}}(\delta t, t)
&&
=\frac{-e\hbar}{2mV}
\sum\limits_{\textbf{k}, \sigma}(2\textbf{k}+\textbf{q}_{\text{pr}})
\left\langle c^{\dagger}_{\textbf{k},\sigma}c^{\phantom{\dagger}}_{\textbf{k}+\textbf{q}_{\text{pr}},\sigma}\right\rangle (\delta t, t) 
\nonumber\\
&&
\; \quad- \frac{e^2}{m V}
\sum\limits_{\textbf{k}, \textbf{q}, \sigma}
{\bf A}_{\textbf{q}_{\text{pr}} - \textbf{q}} \left\langle c^{\dagger}_{\textbf{k},\sigma}c^{\phantom{\dagger}}_{\textbf{k}+\textbf{q} ,\sigma}\right\rangle (\delta t, t)    . \;
\end{eqnarray}
Formally, the current depends on two times, namely the delay time $\delta t$ between pump and probe pulse
and the actual time $t$ at which it is measured or computed, respectively.
For the numerical calculations we neglect the second term in Eq.~\eqref{defCurrent}
 since it only results in  a constant offset of the imaginary part of the conductivity spectra.
Then, the pump-probe conductivity  $\sigma ( \delta t,  \omega )$ is obtained from Eq.~\eqref{defCurrent}  via
\begin{subequations} 
\label{eq:sigma}
\begin{align}
\sigma(\delta t , \omega)=\frac{j(\delta t, \omega)}{i\omega A( \delta t , \omega)},
\end{align}
where 
\begin{eqnarray}
j ( \delta t,  \omega ) =  \int_{-\infty}^{\infty} d t \;  \hat{{\bf e}}_y  \cdot {\bf j}_{{\bf q}_{\textrm{pr}}} ( \delta t, t) \, e^{i \omega t}  
\end{eqnarray}
and 
\begin{eqnarray}
A ( \delta t, \omega ) =  \int_{-\infty}^{\infty} d t \;  \hat{{\bf e}}_y  \cdot  {\bf A}_{{\bf q}_{\textrm{pr}} }
( \delta t, t) \, e^{i \omega t}  
\end{eqnarray}  
\end{subequations}
denote the Fourier transformed $y$~components of the current density ${\bf j}_{{\bf q}_{\textrm{pr}}}(\delta t, t)$ and the vector potential
${\bf A}_{{\bf q}_{\textrm{pr}}} (\delta t, t)$ of the probe pulse, respectively. 
In the literature, there are also other ways discussed to determine the conductivity; see, for instance, 
Ref. \onlinecite{Eckstein2010}. However, we think that in the present context of a pump and a probe pulse
the above procedure suggests itself and is closest to what is experimentally done.
Hence, the pump-probe conductivity is fully determined by the time evolution of the expectation values 
$\langle c^{\dagger}_{\textbf{k},\sigma}c^{\phantom{\dagger}}_{\textbf{k}+\textbf{q}_{\text{pr}},\sigma} \rangle (\delta t, t)$,  which we numerically compute by integrating the corresponding equations of motion. 
To keep the notation light, we in the following omit the dependence on the
delay time $\delta t$ because it is externally fixed by the timing of the pump and the probe pulse.
Below we discuss how the dependence on the time $t$ is computed.

\subsection{Density-Matrix Formalism}
\label{sec:DMF}

Using the density-matrix formalism, we derive  equations of motion for the quasiparticle densities and the mean phonon amplitudes $\langle b^{\phantom{\dagger}}_{\textbf{p}}\rangle$ and $\langle b^{\dagger}_{-\textbf{p}}\rangle$ in this section. 
For this purpose, it is advantageous to
 perform a Bogoliubov transformation of the  electron operators which diagonalizes the Hamiltonian $H_{\text{sc}}$ in the initial state.
That is, we introduce new fermionic operators $\alpha_{\textbf{k}}$ and $\beta_{\textbf{k}}$, with
\begin{align}
\alpha_{\textbf{k}}=u_{\textbf{k}}c^{\phantom{\dagger}}_{\textbf{k},\uparrow}-v_{\textbf{k}}c^{\dagger}_{\textbf{-k},\downarrow}, 
\qquad 
\beta_{\textbf{k}} =v_{\textbf{k}} c^{\dagger}_{\textbf{k},\uparrow}+ u_{\textbf{k}} c^{\phantom{\dagger}}_{-\textbf{k},\downarrow} ,
\end{align}
where 
$
v_{\textbf{k}}= \Delta (t_i ) / |\Delta (t_i ) | \sqrt{ ( 1 - \epsilon_{\textbf{k}} /  E_{\textbf{k}}   )/2 }
$,
$
u_{\textbf{k}}=\sqrt{  ( 1+ \epsilon_{\textbf{k}} /  E_{\textbf{k}}  )/2}
$,
and $E_{\textbf{k}}=\sqrt{\epsilon^2_{\textbf{k}}+|\Delta (t_i ) |^2}$.
We emphasize that the coefficients $u_{\textbf{k}}$ and $v_{\textbf{k}}$ do not depend on time, i.e.,
the temporal evolution of the quasiparticle densities is computed with respect to a fixed time-independent
Bogoliubov-de Gennes basis in which the initial state is diagonal.
All physical observables, such as the electric current density ${\bf j}_{{\bf q}_{\textrm{pr}}} (t)$, the order parameter amplitude $| \Delta (t) |$, and the 
lattice displacement $U ( {\bf r}, t)$ can 
now be expressed in terms of the new Bogoliubov quasiparticle densities $\langle\alpha^{\dagger}_{\textbf{k}}\alpha^{\phantom{\dagger}}_{\textbf{k}' } \rangle$, 
$\langle\beta^{\dagger}_{\textbf{k}}\beta^{\phantom{\dagger}}_{\textbf{k}' } \rangle$,
$\langle\alpha^{\dagger}_{\textbf{k}}\beta^{\dagger}_{\textbf{k}' } \rangle$,
 and $\langle\alpha^{\phantom{\dagger}}_{\textbf{k}}\beta^{\phantom{\dagger}}_{\textbf{k}' } \rangle$. 
For example, for the current density ${\bf j}_{{\bf q}_{\textrm{pr}}} (t)$ we find
\begin{eqnarray}
&& 
{\bf j}_{{\bf q}_{\textrm{pr}}} (t) 
=
\frac{-e\hbar}{2mV}\sum_{\bf k}(2{\bf k}+{\bf q}_{\textrm{pr}})
\\
&&
\; \times
\Big[  
(u_{{\bf k}}v_{{\bf k}+{\bf q}_{\textrm{pr}}}-v_{{\bf k}}u_{{\bf k}+{\bf q}_{\textrm{pr}}})
\big(
\langle\alpha^{\dagger}_{\textbf{k}}\beta^{\dagger}_{{\bf k}+{\bf q}_{\textrm{pr}} } \rangle   
+
\langle \alpha^{\phantom{\dagger}}_{{\bf k}+{\bf q}_{\textrm{pr}} } \beta^{\phantom{\dagger}}_{\textbf{k}} \rangle 
\big)
\nonumber\\
&&
\; +(u_{{\bf k}}u_{{\bf k}+{\bf q}_{\textrm{pr}}}+v_{{\bf k}}v_{{\bf k}+{\bf q}_{\textrm{pr}}}) 
\big(
 \langle\alpha^{\dagger}_{\textbf{k}}\alpha^{\phantom{\dagger}}_{{\bf k}+{\bf q}_{\textrm{pr}} } \rangle 
- 
\langle\beta^{\dagger}_{\textbf{k}}\beta^{\phantom{\dagger}}_{{\bf k}+{\bf q}_{\textrm{pr}} } \rangle  
\big)
\Big]  . \nonumber
\end{eqnarray}

Due to the interaction term $H_{\text{el-ph}}$ in Eq.~\eqref{modHam}, the kinetic equations for the
single-particle density matrices $\langle\alpha^{\dagger}_{\textbf{k}}\alpha^{\phantom{\dagger}}_{\textbf{k}' } \rangle$, 
$\langle\beta^{\dagger}_{\textbf{k}}\beta^{\phantom{\dagger}}_{\textbf{k}' } \rangle$, etc., 
are not closed: Instead, $H_{\text{el-ph}}$ leads to an infinite hierarchy
of equations of higher-order density matrices~\cite{Ross02}. In order to study the generation of coherent phonons,
it is sufficient to break this hierarchy at  first order of the electron-phonon coupling strength $g_{\textrm{ph}}$.   
That is,   phonon-assisted quantities, such as
$\langle\alpha^{\dagger}_{\textbf{k}}\alpha^{\phantom{\dagger}}_{\textbf{k}+\textbf{q}}b^{\phantom{\dagger}}_{\textbf{p}}\rangle$,
are factorized as follows~\cite{Schny11,Herb03}:
\begin{align} \label{factorizeEq}
\langle\alpha^{\dagger}_{\textbf{k}}\alpha^{\phantom{\dagger}}_{\textbf{k}+\textbf{q}}b^{\phantom{\dagger}}_{\textbf{p}}\rangle
=
\langle\alpha^{\dagger}_{\textbf{k}}\alpha^{\phantom{\dagger}}_{\textbf{k}+\textbf{q}}\rangle
\langle b^{\phantom{\dagger}}_{\textbf{p}}\rangle.
\end{align}
Finite expectation values of the mean phonon amplitudes $\langle b^{\phantom{\dagger}}_{\textbf{p}}\rangle$ and $\langle b^{\dagger}_{-\textbf{p}}\rangle$ correspond
to a nonvanishing lattice displacement,
\begin{align}
U(\textbf{r},t)=\sum_{\textbf{p}}\sqrt{\frac{\hbar}{2M\omega_{\text{ph}}N}}
\left(
 \langle b^{\phantom{\dagger}}_{\textbf{p}} \rangle + \langle b^{\dagger}_{-\textbf{p}} \rangle
\right)e^{+ i\textbf{p}\textbf{r}} ,
\end{align} 
where $M$ denotes the reduced mass of the lattice ions. 
We remark that breaking the hierarchy at first order in $g_{\textrm{ph}}$
amounts to neglecting all correlations among quasiparticles and phonons. In particular, relaxation processes
due to quasiparticle-phonon and phonon-phonon scattering are not taken into account. 
Since we focus on time scales shorter than the coherent-phonon and quasiparticle life-times, 
we neglect  all of these higher-order processes,
which are expected to give rise to an exponential damping of the coherent phonon and order parameter oscillations.
 Recent pump-probe experiments on Nb$_{1-x}$Ti$_x$N films~\cite{matsunagaPRL12,matsunagaPRL13} have shown that these relaxation processes occur on time scales of tens of picoseconds, which is much larger than the period of the order parameter oscillations.

By factorizing higher-order density matrices according to Eq.~\eqref{factorizeEq}, 
a closed set of differential equations for the quasiparticle density matrices and the mean phonon amplitudes can be derived 
using Heisenberg's equation of motion.
 A derivation of these differential equations and some other technical details are given
 in Appendix~\ref{sec:eom}.
In Sects.~\ref{sec:result} and~\ref{sec:Results_zwei}, we numerically solve this set of differential equations to obtain the temporal evolution of  
the order parameter amplitude $| \Delta (t) |$, the lattice displacement $U ({\bf r}, t)$,
and the current density ${\bf j}_{{\bf q}_{\textrm{pr}}} (t)$.
From the Fourier transform of the latter quantity, the pump-probe conductivity $\sigma ( \delta t, \omega )$ is readily obtained using Eq.~\eqref{eq:sigma}. In order to study different time-scale regimes, we adjust in the following the pump-pulse width $\tau_{\textrm{p}}$, the phonon energy $\hbar \omega_{\textrm{ph}}$, and the integrated pump-pulse intensity $|\textbf{A}_{\textrm{p}} |^2 \tau_{\textrm{p}}$. For each regime  we determine 
the signatures of the coherent  Cooper pair and phonon dynamics in the pump-probe response.

\begin{figure*}[t]
 \includegraphics[width=\columnwidth]{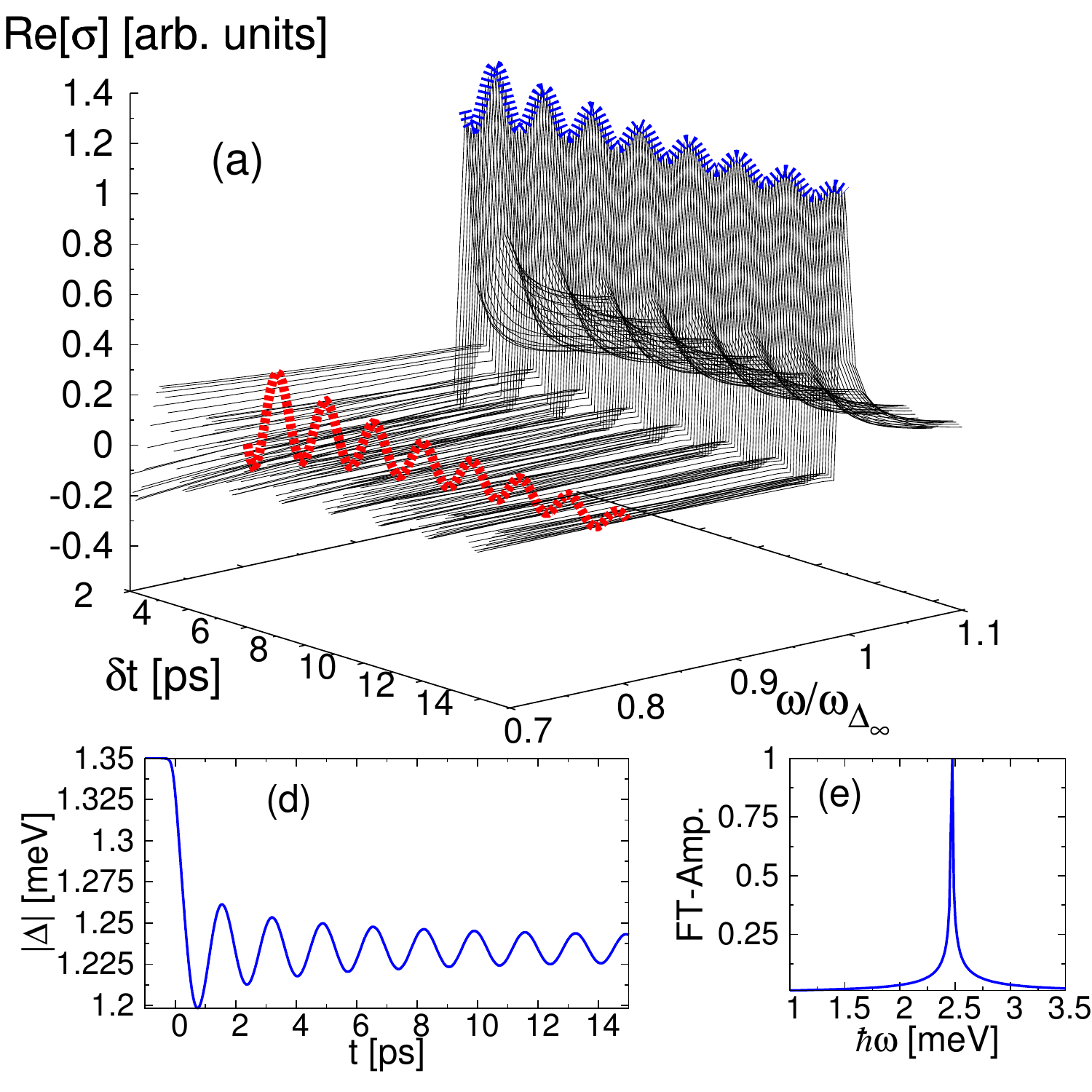}
  \includegraphics[width=\columnwidth]{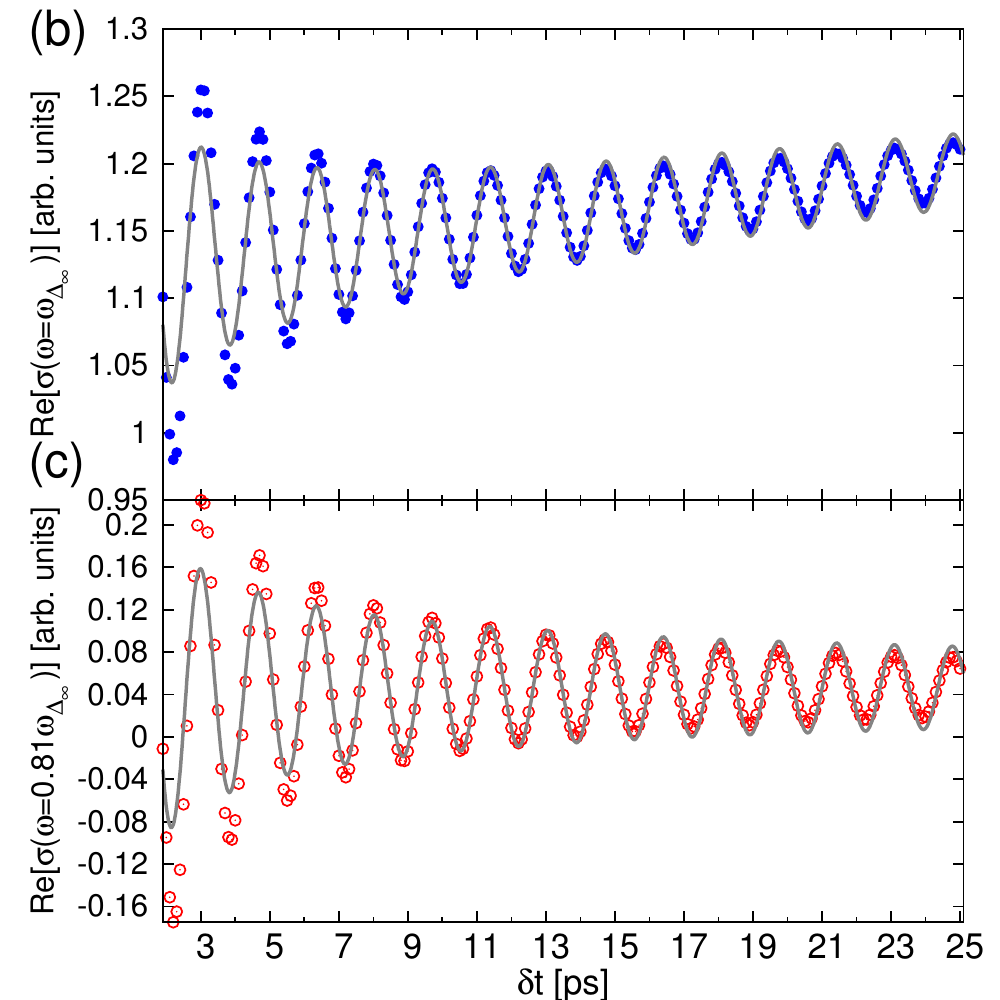}
  \caption{ \label{FigM1} \label{plot:without_phonons}
(Color online) (a) Real part of the pump-probe response, $\mathrm{Re}[ \sigma(\delta t,\omega)]$, versus  $\omega$ and $\delta t >0$
for the nonadiabatic regime [$\tau_{\textrm{p}}=0.5$ ps,  $|\textbf{A}_{\textrm{p}} |=8\cdot10^{-8}$ Js/(Cm)]  in the absence of  phonons. 
(b), (c) Pump-probe signal $\mathrm{Re}[ \sigma(\delta t, \omega) ]$ as a function of delay time $\delta t$ for (b)
$\omega =  \omega_{\Delta_{\infty}} $  and (c) $\omega= 0.81 \, \omega_{\Delta_{\infty}}$.
The gray lines in panels (b) and (c) represent the best fits of Eq.~\eqref{eq:fit_1}
to the numerical data,  see discussion in the text. %
(d),(e)  Temporal evolution of $| \Delta (t) |$ and  spectral distribution of the gap oscillation, respectively, for 
the same parameters as in panel~(a).  } 
\end{figure*}

\section{Pump-probe response in the absence of phonons}\label{sec:result}

Before studying the evolution of the coupled Bogoliubov quasiparticle-phonon system in Sect.~\ref{sec:Results_zwei}, it is instructive to first consider the pump-probe response of a BCS superconductor in the absence of phonons.
Hence, we first solve the set of equations of motion~\eqref{eq:EOMapp} 
for $g_{\textrm{ph}} = 0$ and compute $\sigma ( \delta t, \omega )$  for positive and negative pump-probe delay times $\delta t$.

\subsection{Positive pump-probe delay time}

\begin{figure}[t]
 \includegraphics[width=\columnwidth]{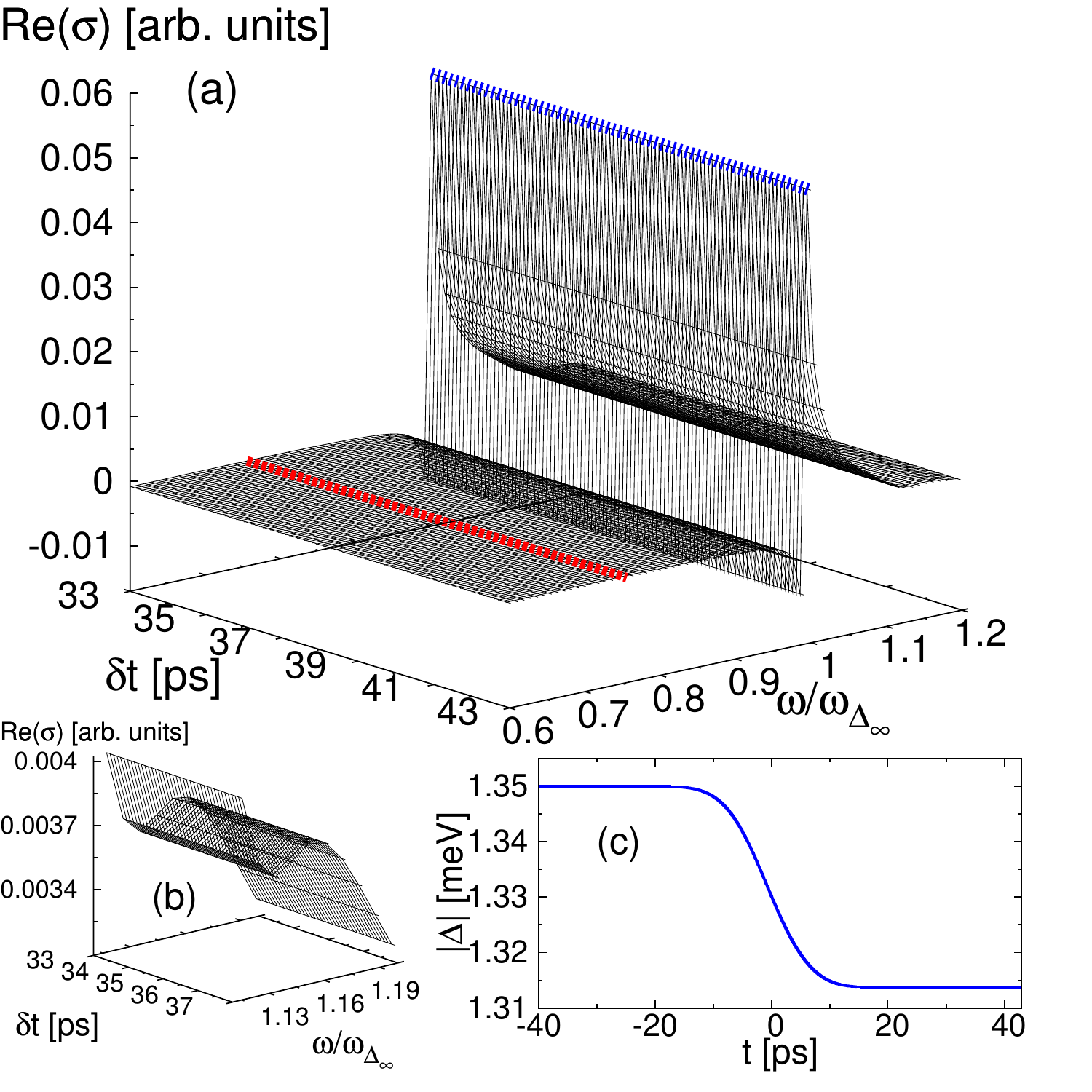}
  \caption{  \label{FigM2} \label{plot:sigma-adi}
(Color online) (a) Real part of the pump-probe  conductivity, $\mathrm{Re}[ \sigma(\delta t,\omega)]$, versus $\omega$ and $\delta t >0$ for the adiabatic regime [$\tau_{\textrm{p}}=20$ ps,  $|\textbf{A}_{\textrm{p}} |=0.5\cdot10^{-8}$ Js/(Cm)]  in the absence of  phonons. 
Panel~(b) shows a detail of the pump-probe response for frequencies just below the pump frequency
$ \omega_{\textrm{p}} = 1.21  \omega_{\Delta_{\infty}}$.  The small dip near $\omega  \approx 1.15 \omega_{\Delta_{\infty}}$ is due to Pauli blocking.
(c) Time dependence of $| \Delta (t) |$ 
for the same parameters as in panel~(a).  } 
\end{figure}

We start by discussing the case where the probe pulse follows the pump pulse after a positive delay time $\delta t$. 
Both the nonadiabatic and the adiabatic regimes are considered, corresponding to $\tau_{\textrm{p}} \ll \tau_{\Delta}$ and
$\tau_{\textrm{p}} \gg \tau_{\Delta}$, respectively; see Figs.~\ref{FigM1} and~\ref{FigM2}.

\subsubsection{Nonadiabatic regime, $\tau_{\textrm{p}}  \ll \tau_{\Delta}$}

In Fig.~\ref{FigM1}(a) we plot  the real part of the pump-probe signal, $\textrm{Re} [ \sigma(\delta t,\omega)]$, versus
delay time $\delta t$ and frequency $\omega$ for the  regime $\tau_{\textrm{p}} \ll \tau_{\Delta}$, where the Bogoliubov quasiparticle densities
build up coherently. The imaginary part of  $\sigma(\delta t,\omega)$, which shows similar features as the real part, is presented in Fig.~\ref{FigM11} of Appendix~\ref{sec:example_im}.
 In the nonadiabatic regime, the ultrafast photoexcitations lead first to a monotonic growth
and then to rapid oscillations in the quasiparticle occupations. Correspondingly, as $t \to \infty$, the order parameter amplitude
$| \Delta (t ) |$ first decreases monotonically and then approaches the asymptotic value $\Delta_{\infty} <  | \Delta ( t_i ) |$
in an oscillatory fashion with oscillation frequency~\cite{volkov1974,amin2004,barankov2004,Yuz05,Yuz06}, 
\begin{eqnarray}
\omega_{\Delta_{\infty}} = 2 \Delta_{\infty} / \hbar ;\   
\end{eqnarray}
 see Figs.~\ref{FigM1}(d) and~\ref{FigM1}(e). 
In the collisionless limit, i.e., in the absence of relaxation processes, the oscillations in the quasiparticle densities are undamped, whereas the
 order parameter oscillations show an algebraic $1 / \sqrt{t}$ decay, due to destructive interference among quasiparticle densities with 
 different momenta [Fig.~\ref{FigM1}(d)]. 
 
Remarkably, we find that  this coherent oscillatory dynamics reveals itself in the pump-probe
 signal through algebraically  decaying oscillations as a function of delay time $\delta t$; see Figs.~\ref{FigM1}(a)-(c).
These oscillations are most prominent at the frequency $\omega_{\Delta_{\infty}}$ corresponding to twice the asymptotic gap energy, i.e., at  $\hbar \omega_{\Delta_{\infty}} = 2 \Delta_{\infty}   =2.4690$ meV, where $\sigma(\delta t,\omega)$ exhibits a sharp edge  as a function of $\omega$; see Figs.~\ref{FigM1}(a) and \ref{FigM1}(b).  
The delay-time dependence of $\textrm{Re} [ \sigma(\delta t,\omega_0 )]$ for fixed  $\omega_0$ is approximately given  by
\begin{align}  \label{eq:fit_1} 
\textrm{Re} [ \sigma(\delta t , \omega_0) ] =A+B\frac{\cos(\omega_{\Delta_{\infty} }\delta t+\Phi)}{\sqrt{\delta t}}+C  \delta t ,
\end{align}
as shown by the excellent fits to the numerical data in Figs.~\ref{FigM1}(b) and~\ref{FigM1}(c). Here, $\Phi$ is an overall phase and $A$, $B$, and $C$ are fit
parameters that depend on $\omega_0$.  Hence, as it is with the order parameter oscillations,  oscillations in the
pump-probe signal are characterized by an amplitude decaying as $1 / \sqrt{ \delta t}$ and a frequency
$\omega_{\Delta_{\infty}} = 2 \Delta_{\infty} / \hbar$ that is determined 
by the asymptotic gap value $\Delta_{\infty} $. We note that the linear increase in 
the pump-probe signal of Fig.~\ref{FigM1}(b) can be attributed to slow oscillations that
are related to the finite size of the system. 

We conclude that the nonadiabatic BCS state dynamics is clearly visible in $\sigma(\delta t,\omega )$ in terms of oscillations
as a function of delay time $\delta t$. This is one of the key findings of the present work. The frequency dependence of the pump-probe signal, on the other hand, does not
show any sign of the order parameter oscillations [Fig.~\ref{FigM1}(a)]~\cite{Papen07}.
 This result is in qualitative agreement 
with recent pump-probe experiments on Nb$_{1-x}$Ti$_x$N thin films~\cite{matsunagaPRL13}.

\subsubsection{Adiabatic regime, $\tau_{\textrm{p}} \gg \tau_{\Delta}$}

It is interesting to contrast the pump-probe response induced by ultrashort pump pulses [Fig.~\ref{FigM1}(a)] with the one in the adiabatic regime, where the pump pulse duration $\tau_{\textrm{p}}$ is much longer than the dynamical time scale of the superconductor 
$\tau_{\Delta}$. The absorption spectrum $\textrm{Re} [ \sigma(\delta t,\omega)]$
as a function of delay time $\delta t$ and frequency $\omega$  for a pump pulse with $\tau_{\textrm{p}} \gg \tau_{\Delta}$ is shown in Fig.~\ref{FigM2}(a). 
 In this adiabatic regime, the pump pulse excites only the normal quasiparticle densities, $\langle\alpha^{\dagger}_{\textbf{k}}\alpha^{\phantom{\dagger}}_{\textbf{k}' } \rangle$ and $\langle\beta^{\dagger}_{\textbf{k}}\beta^{\phantom{\dagger}}_{\textbf{k}' } \rangle$,
 whereas the anomalous ones, $\langle\alpha^{\dagger}_{\textbf{k}}\beta^{\dagger}_{\textbf{k}' } \rangle$ and $\langle\alpha^{\phantom{\dagger}}_{\textbf{k}}\beta^{\phantom{\dagger}}_{\textbf{k}' } \rangle$, remain unoccupied. Hence, the
 order parameter $ \Delta ( t) $ does not oscillate, instead it decreases monotonically towards the asymptotic value $\Delta_{\infty}$  [Fig.~\ref{FigM2}(c)]. Correspondingly, the pump-probe signal  $\sigma(\delta t,\omega)$ does not exhibit any oscillations, neither as a function 
 of delay time  nor of frequency [Fig.~\ref{FigM2}(a)]. As in Fig.~\ref{FigM2}(a), $\textrm{Re} [ \sigma(\delta t,\omega)] $ has a sharp edge
 at the frequency $\omega_{\Delta_{\infty}}$ corresponding to twice the energy of the asymptotic gap value $\Delta_{\infty}$. However,
  otherwise, it is almost featureless, except for a small dip just below the 
 pump frequency $\omega_{\textrm{p}}$; see Fig.~\ref{FigM2}(b). This reduced absorption in the vicinity of $\omega_{\textrm{p}}$ is due to Pauli blocking which leads to a saturation in
 the narrowly peaked quasiparticle  distributions~\cite{Papen07,footnote1}.

\begin{figure}[t]
 \includegraphics[width=\columnwidth]{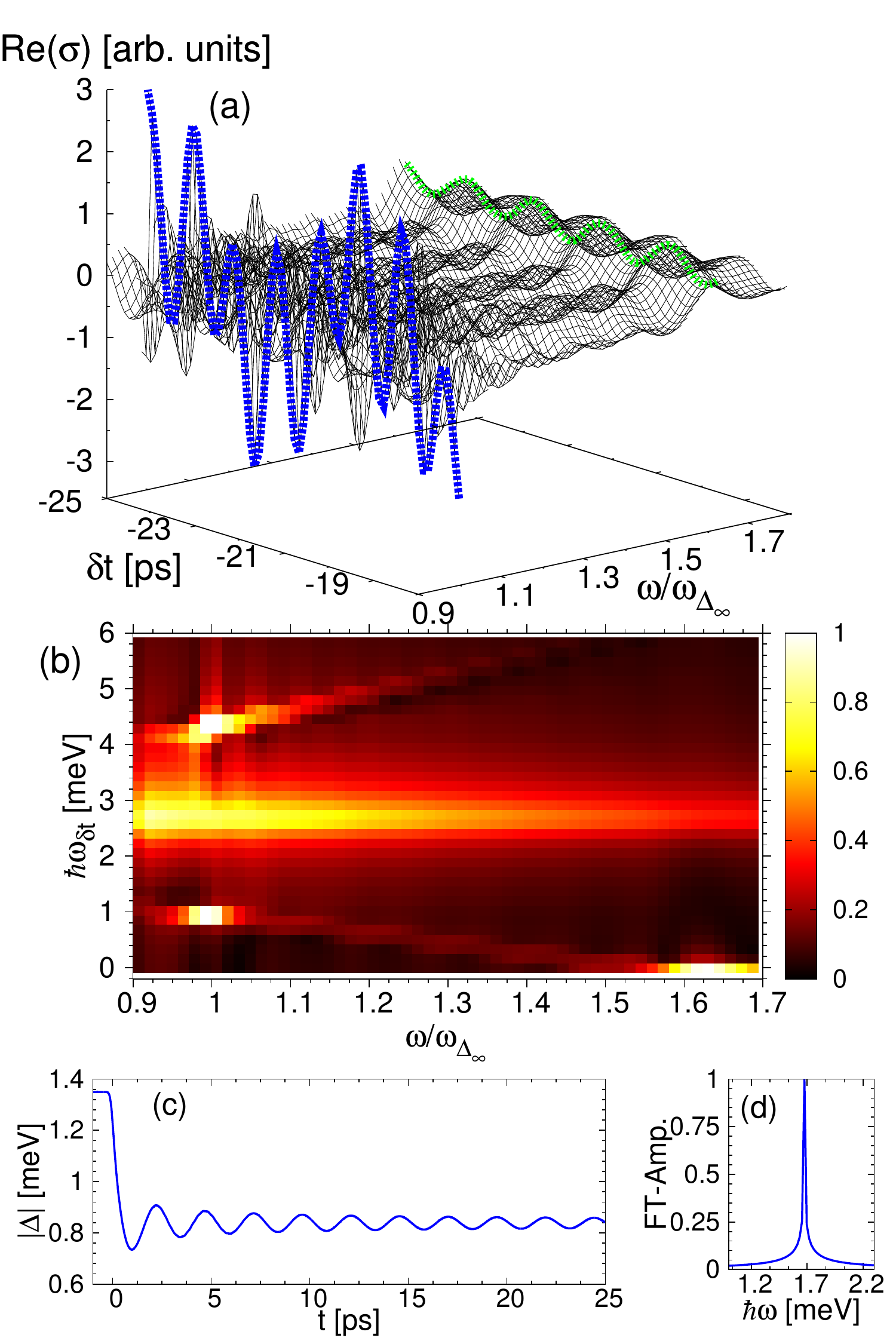}
  \caption{  \label{FigM3} \label{plot:sigma-neg}
  (Color online) (a) Absorption spectrum $\textrm{Re} [ \sigma(\delta t,\omega)]$ versus $\omega$ and $\delta t<0$ for the nonadiabatic regime [$\tau_{\textrm{p}}=0.5$ ps,  $|\textbf{A}_{\textrm{p}} |=13\cdot10^{-8}$ Js/(Cm)] in the absence of  phonons.  (b) Fourier transform of the data in panel (a), i.e.,   $\textrm{Re} ( \sigma ) $ as a function
 of Fourier frequency $\omega_{\delta t}$ and absorption frequency $\omega$.
(c), (d) Time dependence of $| \Delta (t)  |$ and spectral distribution of the gap oscillation, respectively, for the same parameters as in panel~(a).   }
\end{figure}

\begin{figure*}
 \includegraphics[width=\columnwidth]{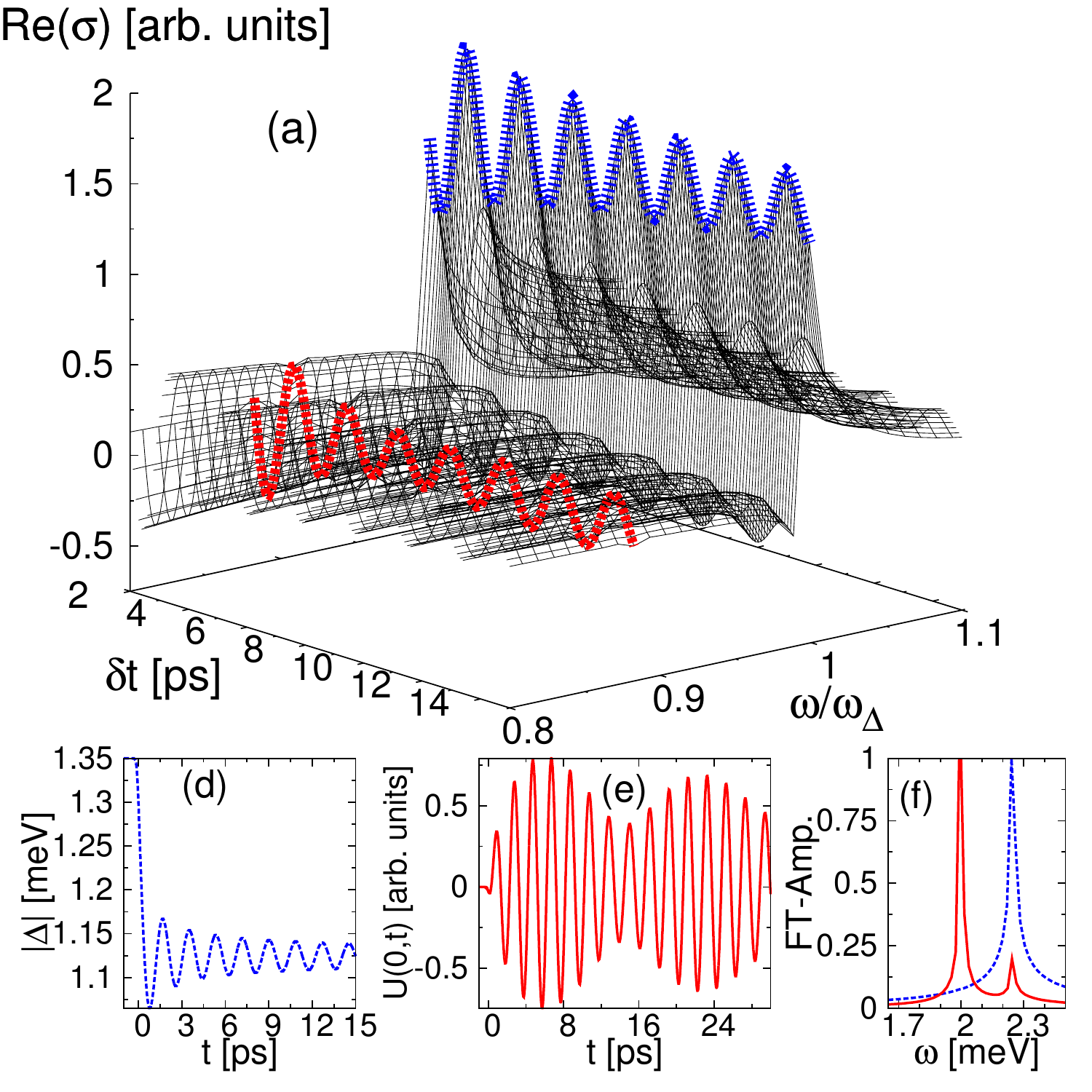}
  \includegraphics[width=\columnwidth]{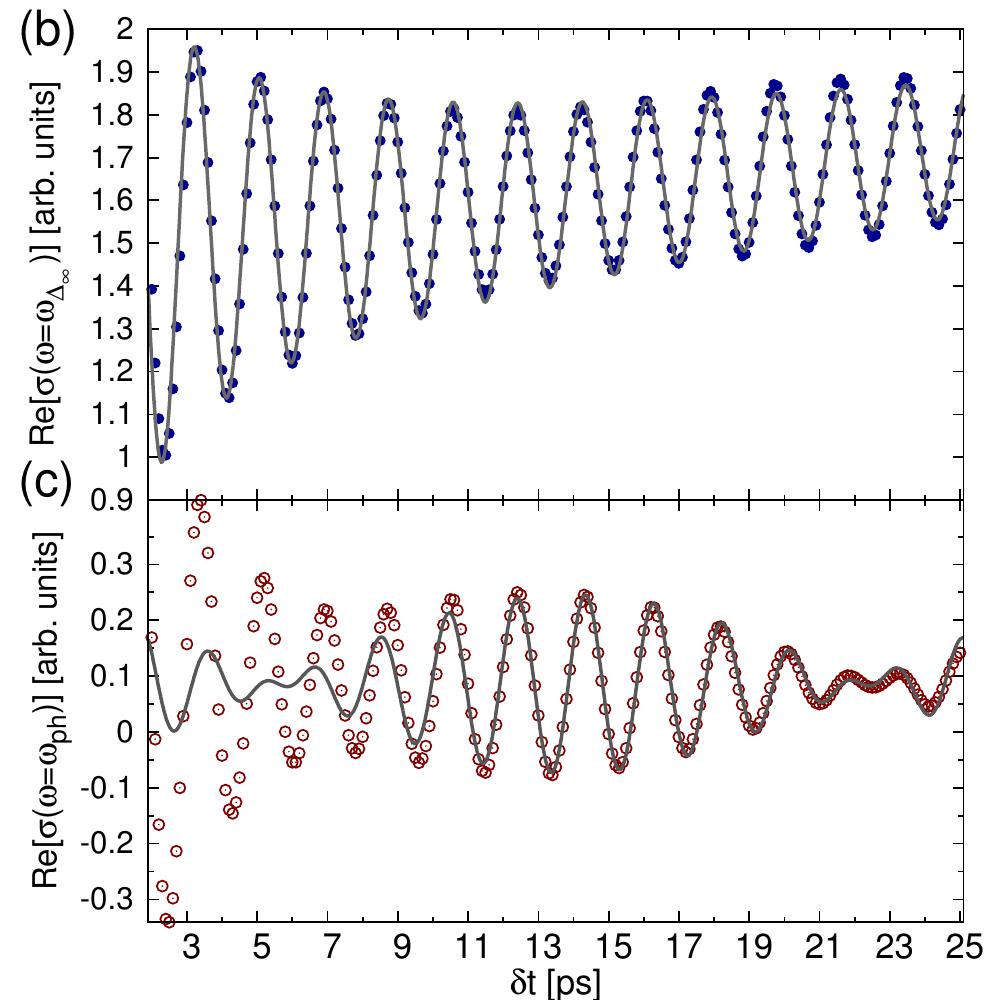}
  \caption{ \label{FigM4} \label{plot:sigma-rg1}
(Color online) 
Real part of the pump-probe signal, $\mathrm{Re} [ \sigma(\delta t,\omega) ]$, versus $\omega$ and $\delta t >0$ for the nonadiabatic regime [$\tau_{\textrm{p}}=0.5$ ps, $|\textbf{A}_{\textrm{p}}|=10\cdot10^{-8}$ Js/(Cm)] in the presence of an optical phonon mode with energy $\hbar \omega_{\textrm{ph}} = 2$~meV and
coupling strength $g_{\text{ph}}=0.1$~meV.  
 (b), (c)  Pump-probe response  $\mathrm{Re} [ \sigma(\delta t,\omega) ]$ as a function of delay time $\delta t$ for (b) $\omega=\omega_{\Delta_{\infty}}$ and (c) $\omega=\omega_{\text{ph}}$.
 The gray lines in panels (b) and (c) represent the best fits of Eqs.~(\ref{eq:fit_1}) and~\eqref{fit_phonon_beating}, respectively, to the numerical data.
(d), (e) Time evolution of the order parameter amplitude $| \Delta (t) |$ and the lattice displacement $U(0,t)$, respectively, for the same parameters as in panel~(a). 
(f) Fourier spectra of the order parameter and coherent phonon oscillations (dashed blue and solid red, respectively).}
\end{figure*}

\subsection{Negative pump-probe delay time}
\label{negativeTimeA}

Let us now turn to the situation where the probe pulse precedes the pump pulse with $\delta t < 0$. 
We focus our analysis on the nonadiabatic case,  
$\tau_{\textrm{p}} \ll \tau_{\Delta}$, 
since in the opposite regime, $\tau_{\textrm{p}} \gg \tau_{\Delta}$, 
the pump-probe signal does not show any interesting characteristics as a function of delay time $\delta t$.

\subsubsection{Nonadiabatic regime, $\tau_{\textrm{p}} \ll \tau_{\Delta}$} 

In Fig.~\ref{FigM3}(a) we present the real part of the pump-probe response, $\textrm{Re} [ \sigma(\delta t , \omega ) ]$,
versus negative delay time $\delta t$ and frequency $\omega$ for an ultrashort pump pulse with 
$\tau_{\textrm{p}} =0.5$~ps $\ll \tau_{\Delta}$.
Both the energy gap before
and after the pump pulse are clearly visible in the frequency dependence of $ \sigma(\delta t , \omega ) $.
That is, the pump-probe signal displays two sharp edges as a function of frequency, one
at twice the gap energy in the initial state $\hbar \omega_{\Delta ( t_i )} =  2 \Delta (t_i)=2.7$ meV and one at  twice the asymptotic gap value
$\hbar \omega_{\Delta_{\infty} } =  2 \Delta_{\infty}=1.677$ meV [green and blue traces in Fig.~\ref{FigM3}(a)]. 
Between those two edges, $ \sigma(\delta t , \omega ) $ shows spectral oscillations in $\omega$ with a frequency $\delta \omega$ that is inversely 
proportional to the delay time, i.e., $\delta\omega =  ( 2 \pi ) / |\delta t| $.\cite{Papen07}  

Interestingly, we find that the pump-probe response $ \sigma(\delta t , \omega ) $ also exhibits a rich oscillatory behavior
 in the delay-time dependence, with multiple frequencies that depend on the absorption energy
 $\hbar \omega$; see Figs.~\ref{FigM3}(a) and~\ref{FigM3}(b). This is revealed most clearly in Fig.~\ref{FigM3}(b),  which shows
 the Fourier transformed pump-probe signal in two-dimensional frequency space, i.e., 
 $\textrm{Re} ( \sigma  )  $ as a function of Fourier frequency $\omega_{\delta t}$ of the delay time 
and absorption frequency $\omega$. 
We observe that for the absorption energy $\hbar \omega$ within the interval  $0 < \hbar \omega < \hbar \omega_{\Delta_{\infty}}$,  
 $\textrm{Re} [ \sigma(\delta t , \omega )]$
 oscillates in $\delta t$ with the frequency $\omega_{\Delta (t_i)}$.
 The oscillatory behavior of
 $\textrm{Re} [ \sigma(\delta t , \omega )]$ in the interval $\hbar \omega_{\Delta_{\infty}} < \hbar \omega <  \hbar \omega_{\Delta (t_i)}$
is even more intriguing, as it shows signatures of how  the gap decreases while the pump pulse acts on the sample [cf.~Fig.~\ref{FigM3}(c)]. 
In other words, it is found that for these absorption energies 
$\textrm{Re} [ \sigma(\delta t , \omega )]$ 
oscillates in $\delta t$ with three frequencies that are approximately given by
$\omega_{\Delta (t_i)}$, $\omega_{\Delta (t_i)} + \omega $,
 and $\omega_{\Delta (t_i)} - \omega $.

\begin{figure*}
 \includegraphics[width=\columnwidth]{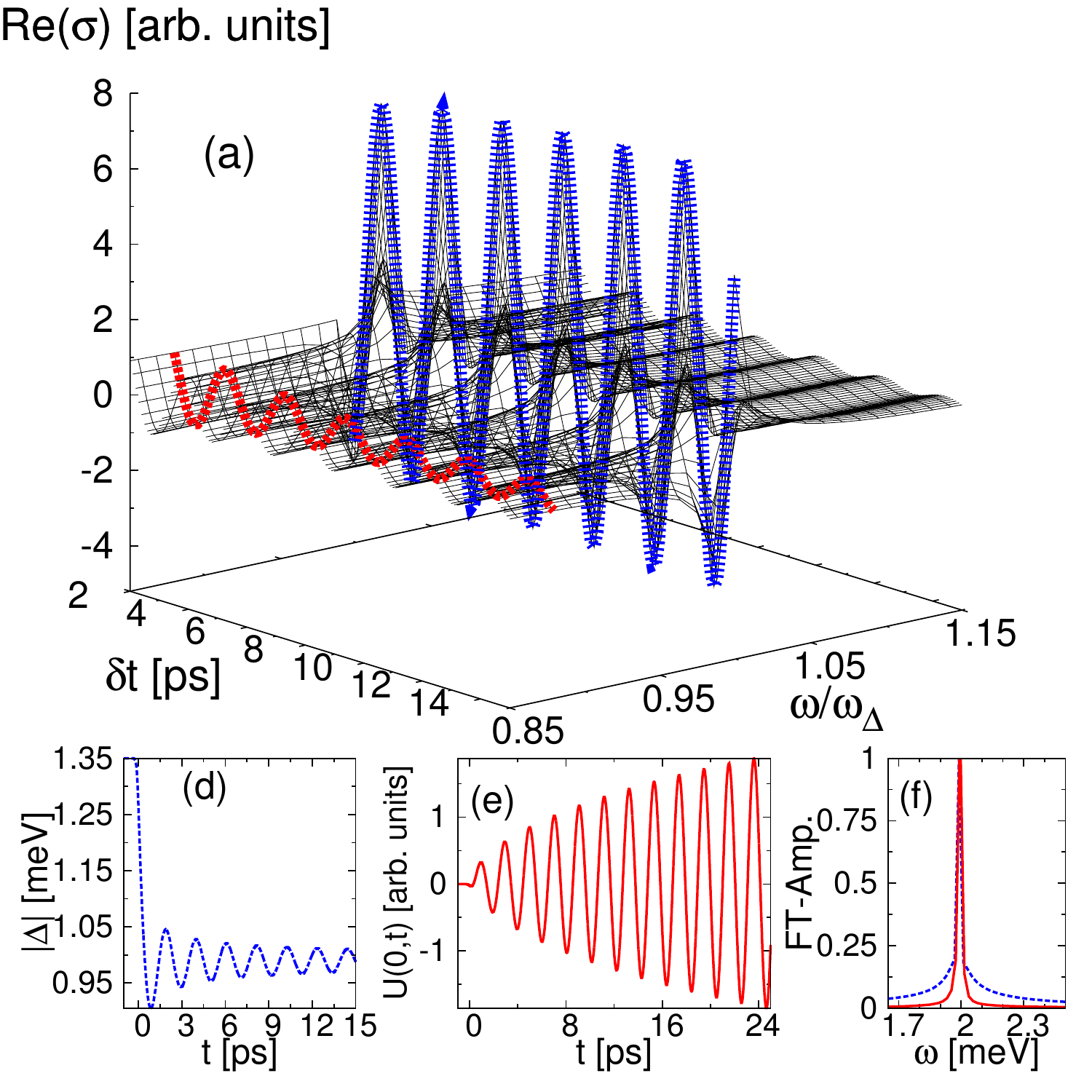}
  \includegraphics[width=\columnwidth]{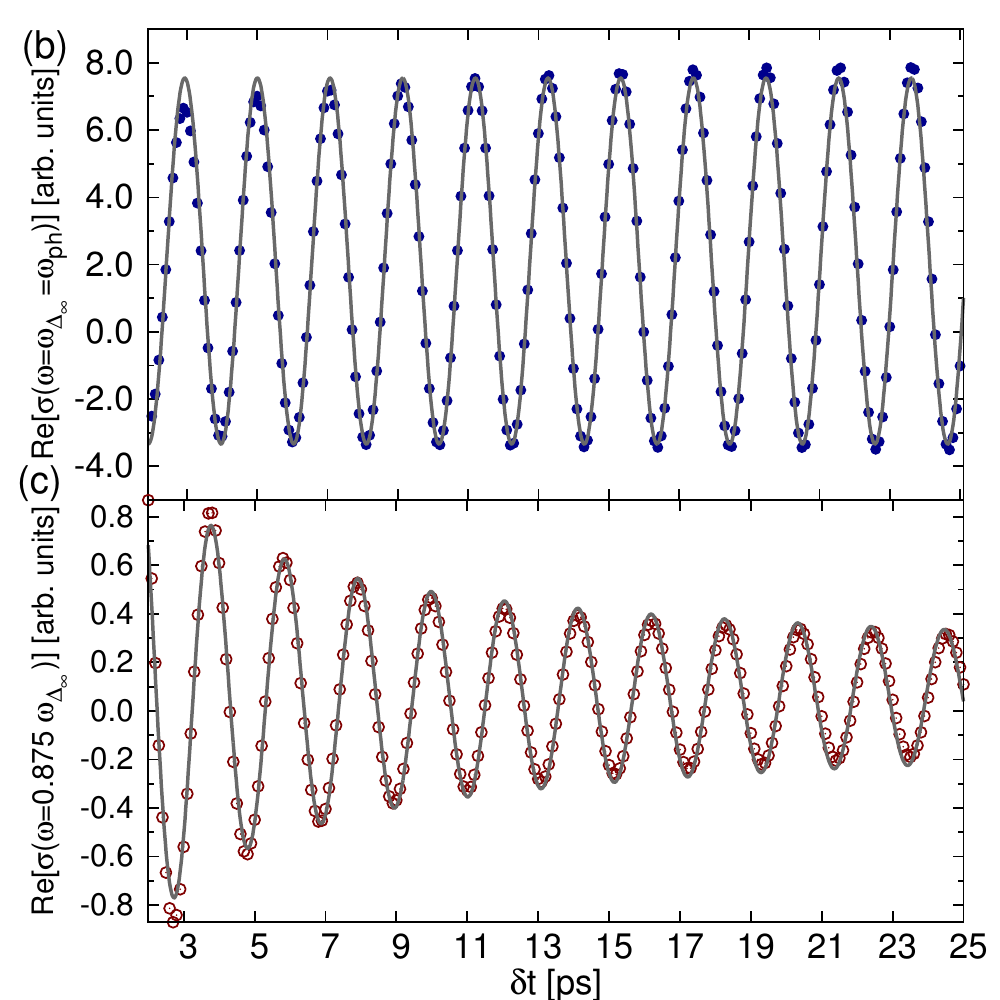}
  \caption{ \label{FigM5} \label{plot:sigma-rg2}
(Color online)  (a) Real part of the the pump-probe response, $\mathrm{Re} [ \sigma(\delta t,\omega)] $, 
versus $\omega$ and $\delta t >0$ 
for the nonadiabatic regime [$\tau_{\textrm{p}}=0.5$ ps, $| {\bf A}_{\textrm{p}} | = 11.64 \cdot 10^{-8}$ Js/(Cm)] in the presence
of an optical phonon mode with energy  at resonance with 
the order parameter oscillations, i.e., $\omega_{\text{ph}} = \omega_{\Delta_\infty} = 2$~meV.
Here the electron-phonon coupling strength is $g_{\text{ph}}=0.1$ meV. 
(b), (c)  Pump-probe signal $\mathrm{Re} [ \sigma(\delta t) ]$ 
as a function of delay time $\delta t$ for (b)  $\omega=\omega_{\Delta_{\infty}}=\omega_{\text{ph}}$ 
and (c) $\omega=0.875\omega_{\Delta_{\infty}}$.
The gray lines in panels (b) and (c) represent the best fits of Eqs.~\eqref{fit_phonon_beating} and~\eqref{eq:fit_1}, respectively, to the numerical data.
(d), (e)  Time dependence of the order parameter amplitude $| \Delta (t) |$ 
and the lattice displacement $U(0,t)$, respectively, for the same parameters as in panel~(a). 
(f) Spectral distribution of the order parameter  and coherent phonon oscillations (dashed blue and solid red, respectively).  }
\end{figure*}

\section{Pump-probe response in the presence of an optical phonon mode}
\label{sec:Results_zwei}

\begin{figure}
 \includegraphics[width=\columnwidth]{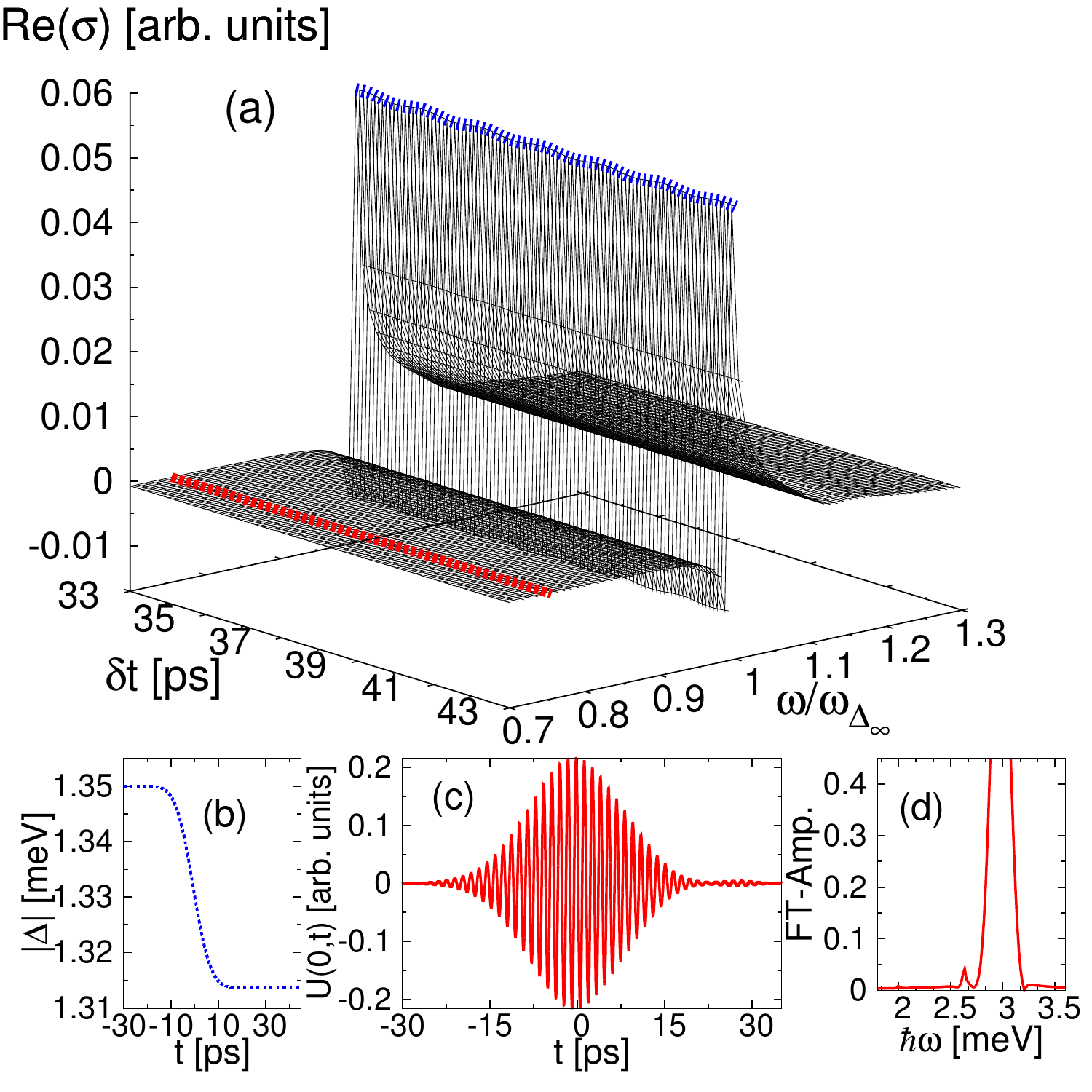}
  \caption{ \label{FigM6} \label{plot:sigma-rg3}
(Color online) (a) Absorption spectrum $\textrm{Re} [ \sigma(\delta t,\omega) ]$ versus $\omega$ and $\delta t > 0$ for the adiabatic regime [$\tau_{\textrm{p}}=20$ ps, $|\textbf{A}_{\textrm{p}}|=0.5\cdot10^{-8}$ Js/(Cm)]
in the presence of an optical phonon mode with energy $\hbar \omega = 2$~meV and coupling strength $g_{\text{ph}}=0.1$~meV. 
(b), (c) Time dependence of the order parameter amplitude $| \Delta (t) |$ and the lattice displacement $U(0,t)$, respectively, for the same parameters as in panel (a). (d)~Spectral distribution of the coherent phonon oscillations. }
\end{figure}

\begin{figure}
 \includegraphics[width=\columnwidth]{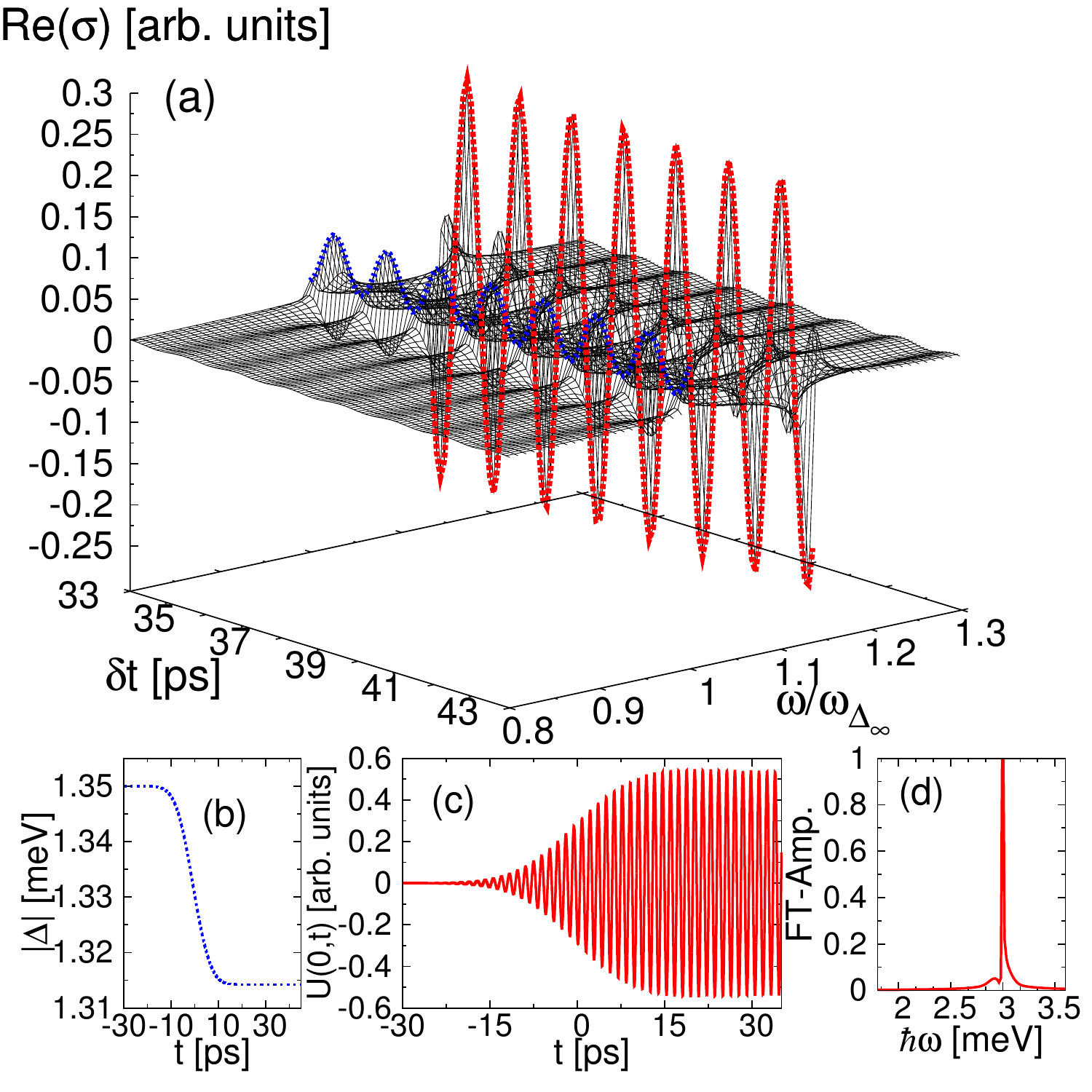}
  \caption{ \label{FigM8a} 
 (Color online) (a) Real part of the pump-probe signal, $\textrm{Re} [ \sigma(\delta t,\omega) ]$, versus $\omega$ and $\delta t > 0$ for the adiabatic regime [$\tau_{\textrm{p}}=20$ ps, $|\textbf{A}_{\textrm{p}}|=0.5\cdot10^{-8}$ Js/(Cm)]
in the presence of an optical phonon mode
 that is in resonance with pump pulse energy, 
i.e., $\hbar \omega_{\textrm{ph}} = \hbar \omega_{\textrm{p}} = 3$~meV.
The electron-phonon coupling strength is taken to be $g_{\text{ph}}=0.1$~meV. 
(b), (c) Time dependence of the order parameter amplitude $| \Delta (t) |$ and the lattice displacement $U(0,t)$, respectively, for the same parameters as in panel (a). (d)~Spectral distribution of the coherent phonon oscillations. }
\end{figure}

\begin{figure}
 \includegraphics[width=\columnwidth]{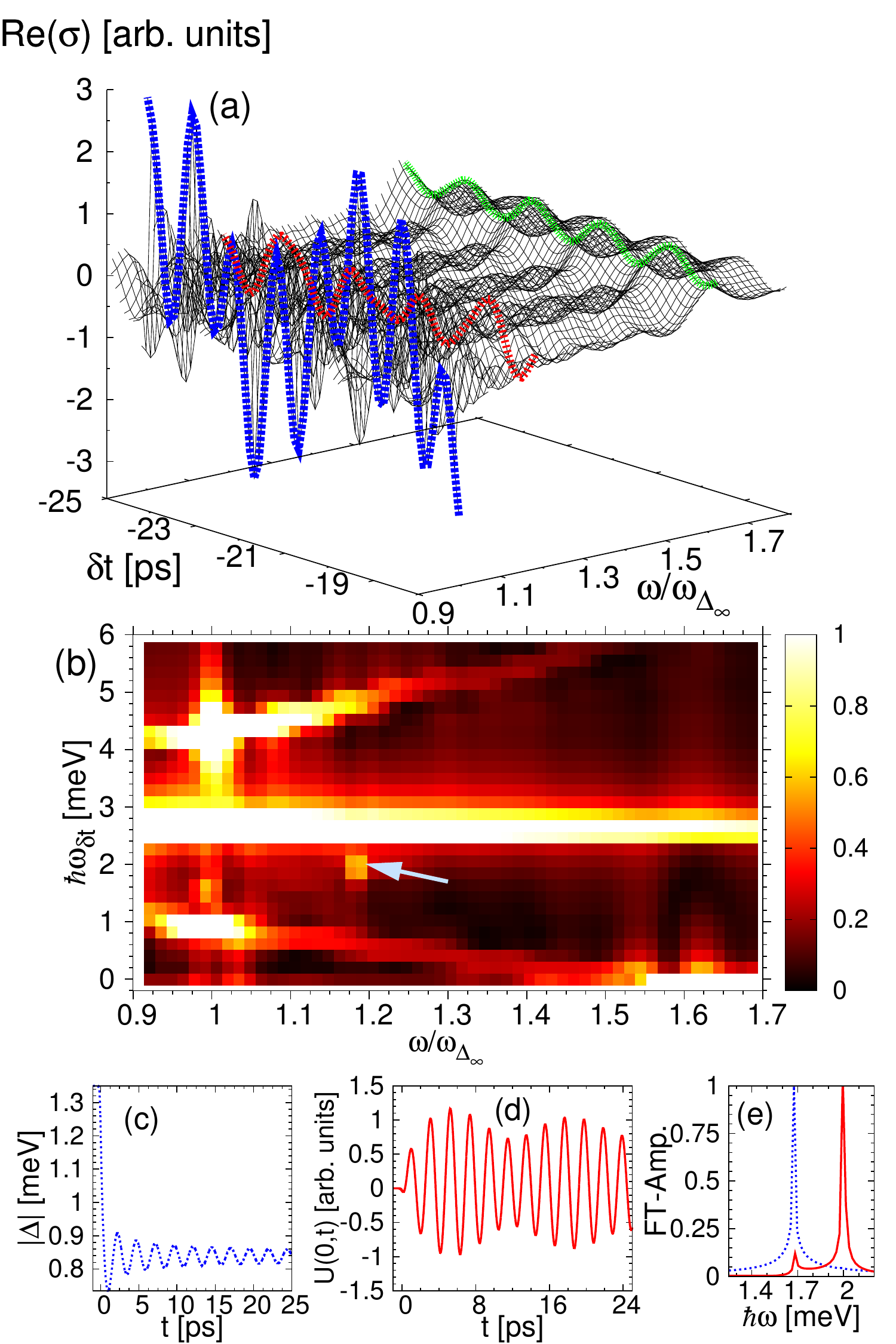}
  \caption{ \label{FigM7} \label{plot:sigma-neg-phonon}
 (Color online) 
(a) Real part of the pump-probe signal, $\textrm{Re} [ \sigma(\delta t,\omega)]$, versus $\omega$  and $\delta t <0$ for the nonadiabatic 
regime [$\tau_{\textrm{p}}=0.5$ ps,  $|\textbf{A}_{\textrm{p}}|=13\cdot10^{-8}$ Js/(Cm)] in the presence of a phonon mode with 
energy $\hbar \omega_{\textrm{ph}} = 2$~meV $ \approx 1.19 \, \hbar \omega_{\Delta_{\infty}}$  and coupling strength $g_{\text{ph}}=0.1$~meV. 
 (b) Fourier transform of the data in panel (a), i.e.,   $\textrm{Re} ( \sigma  )$ as a function of Fourier frequency
$\omega_{\delta t}$ and absorption frequency $\omega$.
(c), (d) Time dependence of the order parameter amplitude $| \Delta (t) |$ and the lattice displacement $U (0, t)$, respectively, 
for the same parameters as in panel (a). 
(e) Spectral distribution of the order parameter and coherent phonon oscillations (dashed blue and solid red lines, respectively). } 
\end{figure}

In this section we analyze the pump-probe response of a superconductor coupled to an optical phonon.
To that end,  we numerically solve the set of kinetic equations~\eqref{eq:EOMapp} for finite electron-phonon coupling $g_{\textrm{ph}}$. 
The correlation expansion used to derive the equations of motion of the coupled  Bogoliubov quasiparticle-phonon system 
is valid as long as $g_{\textrm{ph}}$ is smaller than the superconducting energy scales. Hence, for the numerical calculations we take $g_{\textrm{ph}} = 0.1$~meV $\ll \Delta ( t_i )$, 
in which case the influence of the phonon subsystem on the superconductor becomes negligibly small.
As in the previous section, we focus  on ultrashort pump pulses with $\tau_{\textrm{p}} \ll \tau_{\textrm{ph}}, \tau_{\Delta}$ that drive
both the quasiparticle and the phonon subsystems in a nonadiabatic fashion.

\subsection{Positive pump-probe delay time} 

First, we examine the results for positive pump-probe delay time, where the probe pulse follows the pump
pulse after the delay time $\delta t >0$.

\subsubsection{Nonadiabatic regime, $\tau_{\textrm{p}} \ll \tau_{\textrm{ph}}  \approx  \tau_{\Delta}$}
\label{posTnonAdia}

The most interesting case corresponds to the situation where an ultrashort pump pulse drives quasiparticle oscillations 
with a  frequency that is close to resonance with the phonon energy, i.e., $\tau_{\textrm{p}}  \ll \tau_{\textrm{ph}}  \approx  \tau_{\Delta}$, 
see Figs.~\ref{FigM4}, \ref{FigM5},  and~\ref{FigM12}.  Figure~\ref{FigM4}(a)  
displays the real
part of the pump-probe signal as a function of delay time and frequency for a pump pulse with width $\tau_{\textrm{p}}=0.5$ ps and
 amplitude $| {\bf A}_{\textrm{p} } |  =  10 \cdot 10^{-8}$~Js/(Cm).
For this choice of  pump-pulse intensity, the order parameter oscillations are close to resonance with 
the phonon frequency, i.e.,  $ | 2 \Delta_{\infty } / \hbar   - \omega_{\textrm{ph}} | \ll  \omega_{\textrm{ph}}$ [Fig.~\ref{FigM4}(f)].
This leads to a beating behavior in the lattice displacement $U ( {\bf r}, t)$~\cite{Schny11}; see Fig.~\ref{FigM4}(e).
Correspondingly,  we observe two distinct edges in the frequency dependence of $\sigma ( \delta t, \omega )$, one
at twice the asymptotic gap value  $\hbar\omega_{\Delta_{\infty}} = 2 \Delta_{\infty}  =2.2491$ meV  
and one at the phonon energy $\hbar\omega_{\text{ph}}=2$ meV [blue and red traces in Fig.~\ref{FigM4}(a)].
The maxima of these two edges show an oscillatory behavior as a function of delay time [Figs.~\ref{FigM4}(b) and~\ref{FigM4}(c)],
reflecting the nonadiabatic dynamics of both the phonon mode and the Bogoliubov quasiparticles.

The delay-time evolution of $\textrm{Re} [ \sigma(\delta t, \omega) ]$  at $\omega_{\Delta_{\infty}}$ is well described by Eq.~(\ref{eq:fit_1}), as 
evidenced by the good fit in Fig.~\ref{FigM4}(b). That is, $\textrm{Re} [ \sigma(\delta t , \omega_{\Delta_{\infty}}  ) ]$ oscillates
with frequency $\omega_{\Delta_{\infty}}$ and an amplitude decaying as $1 / \sqrt{ \delta t}$. The oscillations  at 
$\omega_{\textrm{ph}}$, on the other hand, exhibit a beating phenomenon, approximately given by
\begin{align} \label{fit_phonon_beating}
\textrm{Re} [ \sigma(\delta t , \omega_{\textrm{ph}} ) ] 
=
A + B \cos \left(  \omega_+ \delta t + \Phi_1 \right)
\cos \left( \omega_- \delta t  + \Phi_2 \right),
\end{align}
where $\omega_{\pm} = (  \omega_{\Delta_{\infty}} \pm  \omega_{\textrm{ph}} ) / 2$, and $A$, $B$, $\Phi_1$, and $\Phi_2$ are fit parameters. 
As demonstrated by the gray lines in Fig.~\ref{FigM4}(c), Eq.~\eqref{fit_phonon_beating} fits our numerical  results well. 
Just as the lattice displacement $U ( {\bf r}, t)$, $\textrm{Re} [ \sigma(\delta t , \omega_{\textrm{ph}} ) ] $ exhibits
 quantum beats, i.e., it oscillates with frequency $\omega_+$ and an amplitude that is modulated with frequency $\omega_-$. 
Note that the deviations between the fit function \eqref{fit_phonon_beating} and the numerical results of Fig.~\ref{FigM4}(c)  for $\delta t \lesssim 7$~ps are due to a transient oscillatory behavior.

By adjusting the pump-pulse intensity, we can bring the order parameter oscillations into exact resonance with the phonon mode~\cite{Schny11}. 
This is illustrated in Fig.~\ref{FigM5}, where we plot $\textrm{Re} [ \sigma(\delta t , \omega  ) ] $, $| \Delta (t) |$, and 
$U ( {\bf r}, t)$ for a pump pulse with $\tau_{\textrm{p}} = 0.5$~ps and $| {\bf A}_{\textrm{p}} | = 11.64 \cdot 10^{-8}$~Js/(Cm).
In this resonant case, the lattice displacement oscillates with frequency $\omega_{\textrm{ph}} = \omega_{\Delta_{\infty}} = 2$~meV$/\hbar$ and an amplitude that grows like $\sqrt{t}$; see Fig.~\ref{plot:sigma-rg2}(e). 
Concomitantly,  the frequency dependence of the pump-probe signal $\sigma(\delta t, \omega)$ shows just one sharp edge at
$\hbar\omega_{\Delta_{\infty}}=\hbar\omega_{\text{ph}}=2$ meV, whose maximum oscillates as a function of $\delta t$ 
[blue trace in Fig.~\ref{FigM5}(a)]. Remarkably, these oscillations are undamped and their amplitude is considerably larger than 
in the off-resonant case [compare~Fig.~\ref{FigM4}(b) with Fig.~\ref{FigM5}(b)].
In fact, the $\delta t$ dependence of $\textrm{Re} [ \sigma(\delta t , \omega  ) ] $ at $\omega_{\Delta_{\infty}}$ is very well
captured by formula~\eqref{fit_phonon_beating} with $\omega_+ = \omega_{\Delta_{\infty}}$ and $\omega_- =0$,
as demonstrated by the fits in Fig.~\ref{FigM5}(b).
At absorption energies $\hbar \omega$ different from $\hbar \omega_{\Delta_{\infty}}$, however, $\textrm{Re} [ \sigma(\delta t , \omega  ) ]$  shows   algebraically decaying oscillations in $\delta t$ with frequency $\omega_{\Delta_{\infty}}$ and an amplitude decreasing as $1/ \sqrt{ \delta t}$. This is exemplified in Fig.~\ref{FigM5}(c),
which reveals that the delay-time evolution of the pump-probe signal  for $\omega \ne \omega_{\Delta_{\infty}}$ is well described by Eq.~\eqref{eq:fit_1} [gray lines in Fig.~\ref{FigM5}(c)].

\subsubsection{Adiabatic regime, $\tau_{\textrm{p}} \gg \tau_{\Delta}$, $\tau_{\textrm{ph}}$ }

We contrast the results for the nonadiabatic regime [Figs.~\ref{FigM4} and~\ref{FigM5}] with those for the adiabatic case, 
shown in Fig.~\ref{FigM6}(a), where we present the pump-probe signal for a superconductor  photoexcited by a long pump pulse with pulse duration $\tau_{\textrm{p}} \gg \tau_{\Delta}, \tau_{\textrm{ph}}$. We observe that in this regime the lattice displacement does not oscillate with 
the phonon frequency, but exhibits large
transient oscillations with frequency $\omega_{\textrm{p}}$ that occur within the interval $ \approx [- \tau_{\textrm{p}}, + \tau_{\textrm{p}} ]$ during which the pump laser acts on the system [Figs.~\ref{FigM6}(c) and ~\ref{FigM6}(d)]. The probe signal $\textrm{Re} [ \sigma(\delta t , \omega  ) ] $ is almost featureless as a function of delay time $\delta t$ [Figs.~\ref{FigM6}(a)]. It only displays a sharp edge in the frequency dependence at $\hbar \omega_{\Delta_{\infty}} = 2 \Delta_{\infty}$ and a small dip just below the pump pulse frequency $\omega_{\textrm{p}}$ which arises due to Pauli blocking [cf.~Fig.~\ref{FigM2}(a)]. However, since no coherent phonons with frequency $\omega_{\textrm{ph}}$ are being created,
$\textrm{Re} [ \sigma(\delta t , \omega  ) ] $
exhibits no feature at the phonon energy $\hbar \omega_{\textrm{ph}}$ [red trace in Fig.~\ref{FigM6}(a)].
Also the transient oscillations of  Fig.~\ref{FigM6}(c) are not visible in the pump-probe response as a function
of frequency or delay time.

 In closing, we show in Fig.~\ref{FigM8a}  the real part of the pump-probe signal $\textrm{Re} [ \sigma(\delta t , \omega  ) ] $ for 
the special case, where the central energy of the pump pulse   is in resonance  with the phonon energy,
i.e.,  $\hbar \omega_{\textrm{p}} =\hbar \omega_{\textrm{ph}}$.
 Such a resonant pumping of the phonon leads
to undamped coherent phonon oscillations, which persist even after the pump pulse has passed [Fig.~\ref{FigM8a}(c)]~\cite{Schny11}.
This resonant response of the phonon system in turn gives rise to an enhanced oscillatory behavior in the delay-time dependence of the absorption spectrum $\textrm{Re} [ \sigma ( \delta t, \omega) ]$ at  the phonon frequency $\omega_{\textrm{ph}}$  
[red trace in Fig~\ref{FigM8a}(a)].

\subsection{Negative pump-probe delay time}\label{sec:neg_delay}

Finally, we discuss the pump-probe conductivity with negative delay times $\delta t <0$ for a superconductor coupled to 
optical phonons. As in Sect.~\ref{negativeTimeA} we
focus on the nonadiabatic regime $\tau_{\textrm{p}} \ll \tau_{\Delta}, \tau_{\textrm{ph}}$, in which
 order parameter oscillations as well as coherent phonons are generated [Figs.~\ref{FigM7}(c) and \ref{FigM7}(d)].

\subsubsection{Nonadiabatic regime, $\tau_{\textrm{p}} \ll \tau_{\Delta}, \tau_{\textrm{ph}}$}

In Fig.~\ref{FigM7}(a), the real part of the pump-probe signal is plotted  versus negative delay time $\delta t$ and frequency $\omega$ for a short pump pulse with  $\tau_{\textrm{p}} = 0.5$~ps. 
Similar to Fig.~\ref{FigM3}(a), we observe two sharp edges in the frequency dependence of $\textrm{Re} [\sigma ( \delta t , \omega ) ]$,
one at twice the energy of the asymptotic gap value $\hbar \omega_{\Delta_{\infty}} = 1.677$~meV and a smaller one at twice the gap energy of the unperturbed superconductor $\hbar \omega_{\Delta ( t_i )} =  2 \Delta (t_i)=2.7$~meV
[blue and green traces in Fig.~\ref{FigM7}(a)]. In addition, there appears a third feature at the phonon frequency $\hbar \omega_{\textrm{ph}}  \approx 1.19 \hbar \omega_{\Delta_{\infty}}$, indicated by the red trace in Fig.~\ref{FigM7}(a).
As a function of delay time the pump-probe signal shows an  intricate oscillatory behavior, which 
reflects the nonadiabatic dynamics of both the Bogoliubov quasiparticle and phonon subsystems. 
The spectral distribution of these oscillations in $\delta t$ is depicted in Fig.~\ref{FigM7}(b). It is quite similar to the one
in the absence of phonons (cf.~Sect.~\ref{negativeTimeA}). 
For absorption frequencies $\omega$ within the interval $ \omega_{\Delta_{\infty}} <  \omega <   \omega_{\Delta (t_i)}$,
$\textrm{Re} [\sigma ( \delta t , \omega ) ]$ oscillates in general with three different frequencies, approximately
given by  $\omega_{\Delta (t_i)}$, $\omega_{\Delta (t_i)} + \omega $,
 and $\omega_{\Delta (t_i)} - \omega $. Besides these,  the Fourier spectrum of $\textrm{Re} [\sigma ( \delta t , \omega ) ]$ at the absorption
 energy $\omega = \omega_{\textrm{ph}}$ also shows a peak at the phonon energy $\hbar \omega_{\textrm{ph}} = 2$~meV
  [white arrow in Fig.~\ref{FigM7}(b)].

\section{Summary and Conclusions}\label{sec:conclusion}

In this paper we have used the density matrix formalism to simulate the pump-probe response of nonequilibrium superconductors coupled to optical phonons.  Considering different hierarchies of time scales, we have performed systematic calculations of the pump-probe conductivity,
which  allows for a direct comparison of pump-probe experiments  to the theoretical predictions.
For sufficiently short pump pulses the superconductor can be driven into a nonadiabatic regime, which is
characterized by rapid  oscillations in the Bogoliubov quasiparticle densities. In turn, these sudden changes in the quasiparticle occupations lead to order parameter oscillations and the creation of coherent phonons. We have shown that the pump-probe absorption spectrum as a function of  positive and negative delay times  $\delta t$
shows clear signatures of the coherent nonadiabatic  dynamics of both the quasiparticle and the phonon subsystems. 

In particular, for positive delay times, the absorption spectrum at twice the frequency of the asymptotic gap value exhibits algebraically decaying oscillations in $\delta t$ with the same frequency as the order parameter oscillations (Fig.~\ref{FigM1}). 
The coherent dynamics of the phonons reveals itself in terms of a beating phenomenon as a function of $\delta t$  in the absorption spectrum at the phonon frequency (Fig.~\ref{FigM4}). 
Interestingly, this oscillatory response in the probe spectra can be strongly amplified by bringing the frequency of the order parameter oscillations  into resonance with the phonon energy (Fig.~\ref{FigM5}).
 For negative delay times, the pump-probe signal shows an even richer oscillatory response in the delay-time dependence, with multiple frequencies that change as a function of absorption energy (Figs.~\ref{FigM3} and~\ref{FigM7}).  This intricate behavior contains information
on how the superconducting condensate is depleted while the pump pulse acts on the sample.

 Our theoretical findings are qualitatively consistent with  recent pump-probe experiments 
by Matsunaga \textit{et al.}~\cite{matsunagaPRL13}, which have observed order parameter oscillations in
the pump-probe conductivity of Nb$_{1-x}$Ti$_x$N thin films.
 In agreement with our theoretical result, these oscillations
are algebraically damped and their frequency corresponds to twice the asymptotic gap energy.
These experiments have demonstrated that up to five  cycles of order parameter oscillations can be experimentally observed, and that the coherence of the quasiparticles is maintained over a period of up to $10$~ps. 
The theoretical study of relaxation processes due to quasiparticle-quasiparticle and quasiparticle-phonon scattering,
which in the case of Nb$_{1-x}$Ti$_x$N films become dominant at time scales larger than $10$~ps, 
 remains as an important direction for future research.
Pump-probe spectroscopy has the unique capability to resolve coherent oscillations as well as relaxation processes in the time domain, thereby yielding important information on the intrinsic time and energy scales of the superconductor.\cite{matsunagaPRL12,matsunagaPRL13,mansartPNAS13} We hope that the findings of this paper will stimulate further time-resolved measurements of superconductors in the nonadiabatic regime.

\begin{acknowledgments}
We gratefully acknowledge many useful discussions with A.\ Avella, A.\ Cavalleri, I.\ Eremin,  and C.\ Giannetti.
H.K.\ and G.S.U.\  acknowledge financial support by the Mercator Stiftung. H.K.\
thanks the Max-Planck-Institut FKF Stuttgart for its hospitality. 
\end{acknowledgments}

\appendix

\section{Numerical simulation of\\
 the pump-probe response}
\label{sec:eom}

 In this appendix, we present some technical details about the numerical implementation of the pump-probe response simulation. The closed set
of equations of motion describing the time evolution of a superconductor coupled to a phonon branch is presented in Sec.~\ref{sec:EOM}. The discretization of the Brillouin zone  is discussed in Sect.~\ref{sec:dis} of this appendix. Some further details about the implementation of the probe pulse can be found in Sect.~\ref{sec:probe_technical}. 

\subsection{Equations of motion}\label{sec:EOM}
The equation of motion for the normal quasiparticle density $\langle\alpha^{\dagger}_{\textbf{k}}\alpha^{\phantom{\dagger}}_{\textbf{k}+\textbf{q}}\rangle$ is given by
\begin{subequations} \label{eq:EOMapp}
\begin{widetext}
\begin{eqnarray}\label{eq:eom}
&& i\hbar\frac{\text{d}}{\text{d}t}\left\langle\alpha^{\dagger}_{\textbf{k}}\alpha^{\phantom{\dagger}}_{\textbf{k}'}\right\rangle =
			      (R_{\textbf{k}'}-R_{\textbf{k}})\left\langle\alpha^{\dagger}_{\textbf{k}}\alpha^{\phantom{\dagger}}_{\textbf{k}'}\right\rangle
			      +C_{\textbf{k}'}\left\langle\alpha^{\dagger}_{\textbf{k}}\beta^{\dagger}_{\textbf{k}'}\right\rangle
			      +C^{*}_{\textbf{k}}\left\langle\alpha^{\phantom{\dagger}}_{\textbf{k}'}\beta^{\phantom{\dagger}}_{\textbf{k}}\right\rangle\\
			     & &+\frac{e\hbar}{2m}\sum_{\textbf{q}'=\pm \textbf{q}_0} 2\textbf{k}\textbf{A}_{\textbf{q}'}\left(-L^{+}_{\textbf{k},\textbf{q}'}\left\langle\alpha^{\dagger}_{\textbf{k}+\textbf{q}'}\alpha^{\phantom{\dagger}}_{\textbf{k}'}\right\rangle
			     +L^{+}_{\textbf{k}',-\textbf{q}'}\left\langle\alpha^{\dagger}_{\textbf{k}}\alpha^{\phantom{\dagger}}_{\textbf{k}'-\textbf{q}'}\right\rangle 
			     -M^{-}_{\textbf{k}',-\textbf{q}'}\left\langle\alpha^{\dagger}_{\textbf{k}}\beta^{\dagger}_{\textbf{k}'-\textbf{q}'}\right\rangle\notag
			     -M^{-}_{\textbf{k},\textbf{q}'}\left\langle\alpha^{\phantom{\dagger}}_{\textbf{k}'}\beta^{\phantom{\dagger}}_{\textbf{k}+\textbf{q}'}\right\rangle\right)\notag \nonumber\\
			     & &+\frac{e^2}{2m}\sum_{\textbf{q}'=0,\pm2\textbf{q}_0}\left(\sum_{\textbf{q}_i=\pm \textbf{q}_0} A_{\textbf{q}'-\textbf{q}_i}A_{\textbf{q}_i}\right)\left(-L^{-}_{\textbf{k},\textbf{q}'}\left\langle\alpha^{\dagger}_{\textbf{k}+\textbf{q}'}\alpha^{\phantom{\dagger}}_{\textbf{k}'}\right\rangle
			     +L^{-}_{\textbf{k}',\textbf{q}'}\left\langle\alpha^{\dagger}_{\textbf{k}}\alpha^{\phantom{\dagger}}_{\textbf{k}'-\textbf{q}'}\right\rangle +M^{+}_{\textbf{k}',-\textbf{q}'}\left\langle\alpha^{\dagger}_{\textbf{k}}\beta^{\dagger}_{\textbf{k}'-\textbf{q}'}\right\rangle
			     +M^{+}_{\textbf{k},\textbf{q}'}\left\langle\alpha^{\phantom{\dagger}}_{\textbf{k}'}\beta^{\phantom{\dagger}}_{\textbf{k}+\textbf{q}'}\right\rangle\right)\notag\\
			     & &+\frac{g_{\text{ph}}}{\sqrt{N}}\sum_{\textbf{p}}\left(\left\langle b^{\dagger}_{-\textbf{p}}\right\rangle+\left\langle b^{\phantom{\dagger}}_{\textbf{p}}\right\rangle\right)\left(-L^{-}_{\textbf{k},\textbf{p}}\left\langle\alpha^{\dagger}_{\textbf{k}+\textbf{p}}\alpha^{\phantom{\dagger}}_{\textbf{k}'}\right\rangle
			     +L^{-}_{\textbf{k}',-\textbf{p}}\left\langle\alpha^{\dagger}_{\textbf{k}}\alpha^{\phantom{\dagger}}_{\textbf{k}'-\textbf{p}}\right\rangle  
			     +M^{+}_{\textbf{k},\textbf{p}}\left\langle\alpha^{\phantom{\dagger}}_{\textbf{k}'}\beta^{\phantom{\dagger}}_{\textbf{k}+\textbf{p}}\right\rangle
			     +M^{+}_{\textbf{k}',-\textbf{p}}\left\langle\alpha^{\dagger}_{\textbf{k}}\beta^{\dagger}_{\textbf{k}'-\textbf{p}}\right\rangle\right),  
			     \nonumber
\end{eqnarray}
where 
\begin{align}
R_{\textbf{k}}&=\frac{\epsilon^2_{\textbf{k}}+\text{Re}(\Delta^{*}\Delta_0)}{E_{\textbf{k}}}, 
\qquad
C_{\textbf{k}}=\Delta^{*}_0\left(\frac{\epsilon_{\textbf{k}}}{E_{\textbf{k}}}\left(1-\text{Re}\left(\frac{\Delta}{\Delta_0}\right)\right)-i\,\text{Im}\left(\frac{\Delta}{\Delta_0}\right)\right),
\end{align}
 and with the shorthand notation
$L^{\pm}_{\textbf{k},\textbf{q}} = u_{\textbf{k}}u_{\textbf{k}+\textbf{q}}\pm v_{\textbf{k}}v_{\textbf{k}+\textbf{q}}$ and
$M^{\pm}_{\textbf{k},\textbf{q}} = v_{\textbf{k}}u_{\textbf{k}+\textbf{q}}\pm v_{\textbf{k}+\textbf{q}}u_{\textbf{k}}$.
The equations of motion for the other quasiparticle densities are of similar form\cite{Schny11} and are not presented for brevity.
The equation of motion for the mean phonon amplitude $ \left\langle b_{\textbf{p}}\right\rangle$ is given by
\begin{align} \label{phonEOma}
i\hbar\abl \left\langle b_{\textbf{p}}\right\rangle&=\hbar\omega_{\text{ph}}\left\langle b_{\textbf{p}}\right\rangle
	      +\frac{g_{\text{ph}}}{\sqrt{N}}\sum_{\textbf{k}}\left(L^{-}_{\textbf{k},-\textbf{p}}
								      \left(\left\langle\alpha^{\dagger}_{\textbf{k}-\textbf{p}}\alpha^{\phantom{\dagger}}_{\textbf{k}}\right\rangle
								      +\left\langle\beta^{\dagger}_{\textbf{k}}\beta^{\phantom{\dagger}}_{\textbf{k}-\textbf{p}}\right\rangle\right)
						+M^{+}_{\textbf{k},-\textbf{p}}\left(
								      \left\langle\alpha^{\dagger}_{\textbf{k}-\textbf{p}}\beta^{\dagger}_{\textbf{k}}\right\rangle
								      -\left\langle\alpha^{\phantom{\dagger}}_{\textbf{k}}\beta^{\phantom{\dagger}}_{\textbf{k}-\textbf{p}}\right\rangle\right)
								       \right) .
\end{align}
 Equations~\eqref{eq:eom}-\eqref{phonEOma} together with the equations of motion for the other three quasiparticle densities form a closed
set of differential equations, which we solve using a standard Runge Kutta algorithm.
\end{widetext}
\end{subequations}

\subsection{Discretization}\label{sec:dis}

\begin{figure}
 \includegraphics[width=0.8\columnwidth]{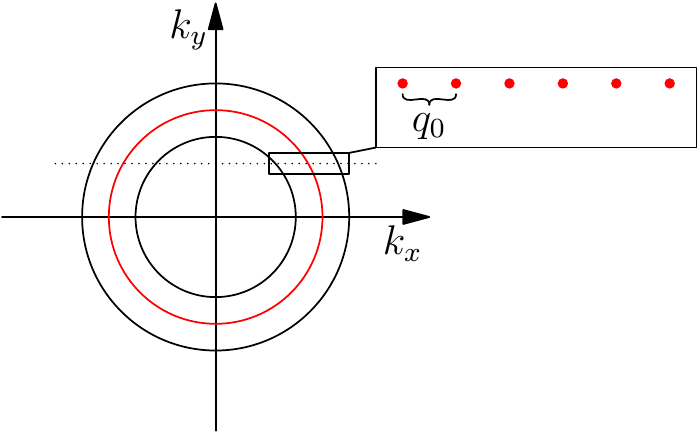}
  \caption{(Color online) Sketch of the  two-dimensional momentum space. The red line represents  the Fermi surface $k=k_{\text{F}}$ and the solid black lines are the boundary given by the radial component $k=k_{\text{F}}\pm k_{\text{c}}$  ($k_{\text{c}}$, radial component of the cutoff vector). The chosen $k_y$ values are indicated by the  dotted line. In the inset the discretization of the $k_x$ values is shown. Only $k_x$ values with a distance of $q_0$  couple to each other.}
  \label{plot:dis}
\end{figure}

To solve the closed set of differential equations for the expectation values, e.g. $\left\langle\alpha^{\dagger}_{\textbf{k}}\alpha^{\phantom{\dagger}}_{\textbf{k}+\textbf{q}}\right\rangle$, numerically, we have to restrict the numbers of considered points in momentum space. The first restriction is that we only take expectation values with indices $\textbf{k}$ and $\textbf{k}+\textbf{q}\in W$ into account. Furthermore, only expectation values with indices ($\textbf{k}$, $\textbf{k}+n\textbf{q}_0$) with an integer $n$ have to be consider, as we can see from Eq.~(\ref{eq:eom}).  The external electromagnetic field breaks translational invariance and may add or subtract momentum $\textbf{q}_0$. For small amplitudes $|\textbf{A}_{\textrm{p}}|$ the off-diagonal elements decrease rapidly as increases because the contribution at $(\textbf{k}$, $\textbf{k}+n\textbf{q}_0)=O(|\textbf{A}_{\textrm{p}}|^{|n|})$. 
Thus, we set all entries with $n>4$ to zero. With this choice of the $\textbf{k}$, $\textbf{k}+\textbf{q}$ values, we are able to solve the equations of motion.
To reduce the numerical effort we perform a quasi-one-dimensional calculation.  The discretization mesh studied is depicted in Fig.~\ref{plot:dis}. 
It has been shown that this quasi-one-dimensional simulation yields a good approximation for two- and three-dimensional systems~\cite{Papen07}.

\subsection{Implementation of pump and probe pulses}
\label{sec:probe_technical}

\begin{figure}
\includegraphics[width=\columnwidth]{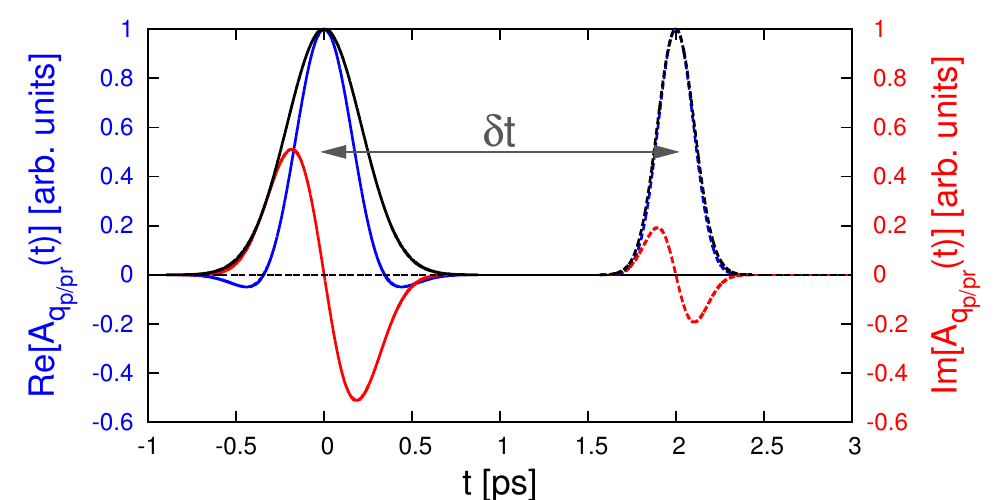}
\caption{ \label{FigM9_new}
 (Color online)  Temporal evolution of the pump field $A_{{\bf q}_{\textrm{p}}} (t)$ (solid lines) and 
and probe field $A_{{\bf q}_\textrm{pr}}$(t) (dashed lines) with pulse widths $\tau_{\textrm{p}}=0.5~\text{ps}$
and $\tau_{\textrm{pr}}=0.25~\text{ps}$, respectively. The black solid (dashed) line represents
the Gaussian envelope of the pump (probe) pulse. The blue and red traces show the real and imaginary 
parts of the pulse fields, respectively. The pump pulse is centered at $t=0~\text{ps}$ and the probe pulse at $t=\delta t = 2$~ps. }
\end{figure}

 We have considered both long and short pump pulses, with widths $\tau_{\textrm{p}} \ll \tau_{\Delta}$ and $\tau_{\textrm{p}} \gg \tau_{\Delta}$, respectively. The width of the probe pulse, however, is always chosen to be short, with $\tau_{\textrm{pr}}=0.25$~ps, such that the nonequilibrium superconductor is probed with a broad range of frequencies. Both the pump and the probe  pulses are cut off at a thousandth of their maximum amplitude. Hence, for $| \delta t | \gtrsim 2 ( \tau_{\textrm{p}} + \tau_{\textrm{pr}} )$ there is no overlap between  pump and probe pulses.
Figure~\ref{FigM9_new} depicts the temporal evolution of the pump and probe pulses.

 For the pump pulse, which has a high intensity, we take into account both linear and nonlinear couplings to the superconductor [cf.~Eq.~\eqref{eq:eom}]. The probe pulse, on the other hand, has weak intensity and is therefore only treated within linear approximation,
i.e., terms of second order and higher in the probe field $A_{\textrm{pr}}(t)$ are neglected.  As a result, the equations of motion simplify. 

To compute the effects of the probe pulse we use further following approximations. First, all off-diagonal terms, such as $\EW{\crea{\alpha}{\kvec}\crea{\beta}{\kvec+\qvec_{\text{pr}}}}$, are zero before the probe pulse is switched on. Second, we use the same momentum grid for pump and probe pulse. Strictly speaking, due to the different wave vectors of the pump and probe pulses, both pulses act on   different momentum grids. However, due  to the small values of the photon wave vector this approximation is valid. For the diagonal elements we approximate 
\begin{align}
\EW{\crea{\alpha}{\kvec+\qvec_{\text{pr}}}\crea{\beta}{\kvec+\qvec_{\text{pr}}}}\approx\EW{\crea{\alpha}{\kvec+\qvec_{\text{p}}}\crea{\beta}{\kvec+\qvec_{\text{p}}}}\, \text{etc.}
\end{align}
and restrict the off-diagonal elements to 
\begin{align}
\EW{\crea{\alpha}{\kvec+m\qvec_{\text{pr}}}\crea{\beta}{\kvec+n\qvec_{\text{pr}}}}=0\,\quad\text{if}\; \; |m-n|>1. \label{eq:approx_non_dia}
\end{align}
 As mentioned above the probe pulse is computed only in linear order in  $A_{\textrm{pr}} (t)$. So we neglect the parts in the equation which are proportional to $\frac{e^2}{2m}$. Additionally, all contributions in linear order of  $A_{\textrm{pr}} (t)$ and  the ones relevant to $g_{\text{ph}}$ simplify due to the approximation Eq. (\ref{eq:approx_non_dia}), because not every addend  in Eq.~(\ref{eq:eom}) has to  be taken into account due to the fact that the corresponding expectation value is set to zero.

\begin{figure}[thp]
 \includegraphics[width=\columnwidth]{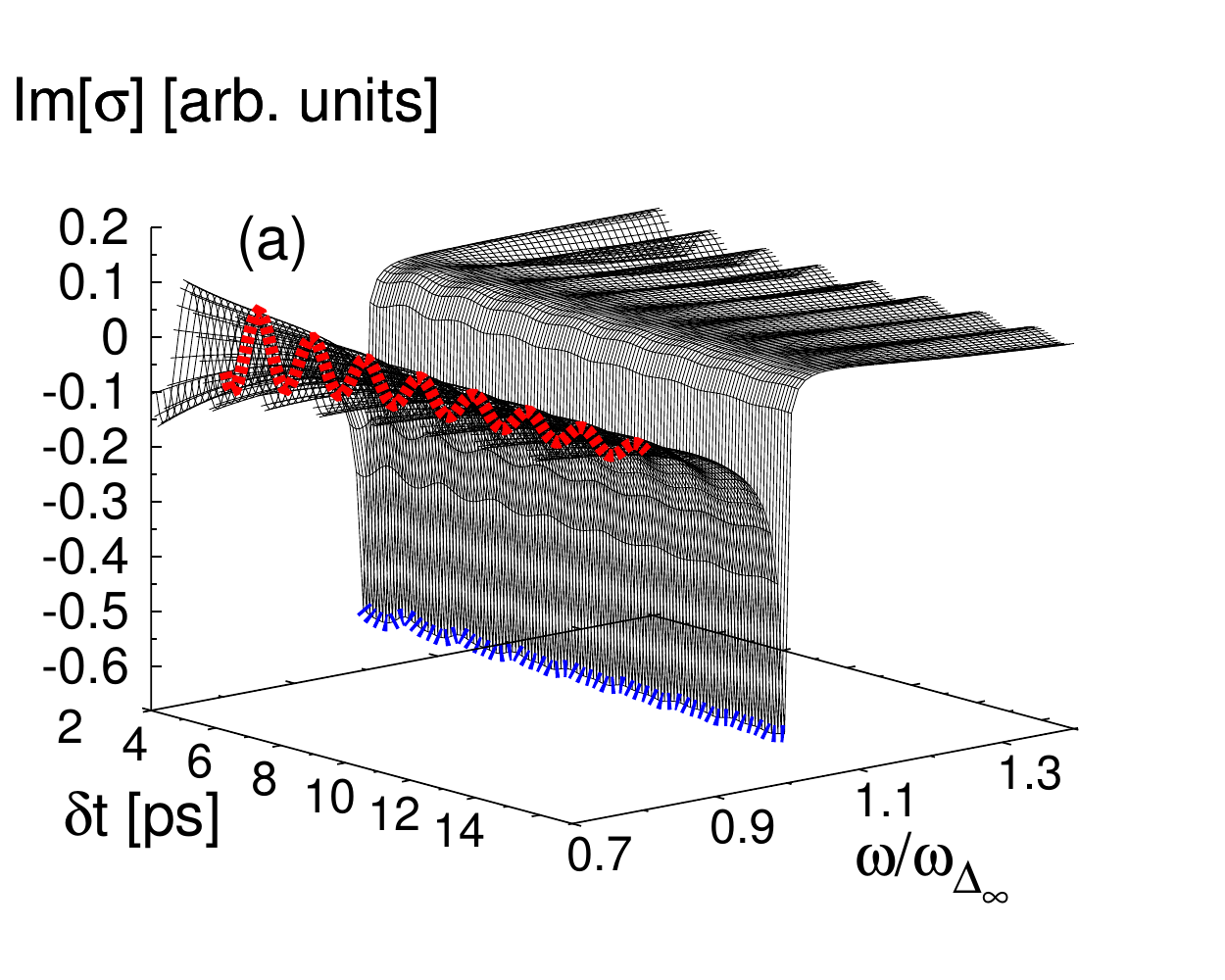}
  \includegraphics[width=\columnwidth]{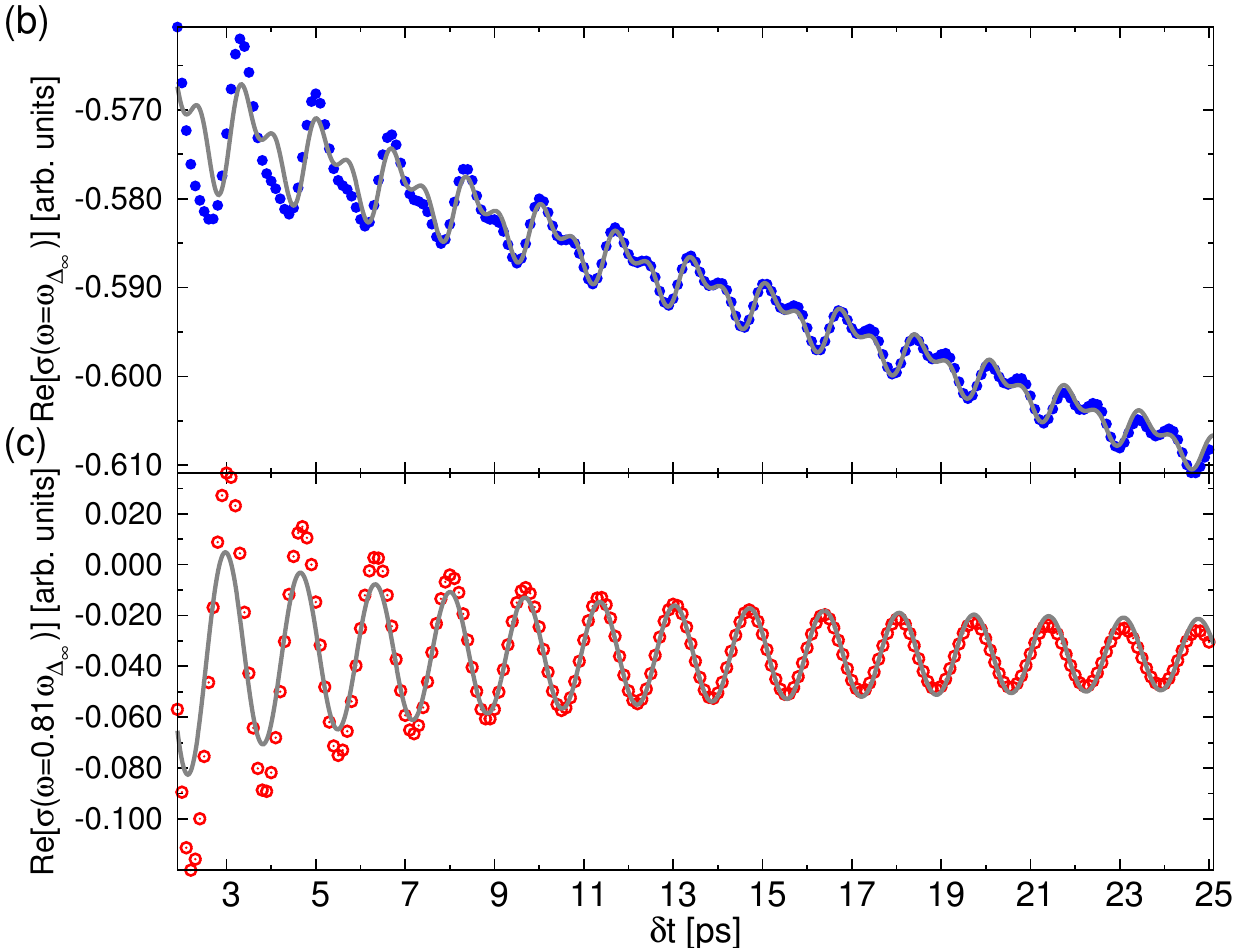}
  \caption{ \label{FigM11} \label{plot:sigma-im_gp_0}
   (Color online) (a) Imaginary part of the pump-probe response, $\mathrm{Im}[ \sigma(\delta t,\omega)]$, versus  $\omega$ and $\delta t >0$
for the nonadiabatic regime [$\tau_{\textrm{p}}=0.5$ ps,  $|\textbf{A}_{\textrm{p}} |=8\cdot10^{-8}$ Js/(Cm)]  in the absence of  phonons. 
(b), (c) Pump-probe signal $\mathrm{Im}[ \sigma(\delta t, \omega) ]$ as a function of delay time $\delta t$ for (b)
$\omega =  \omega_{\Delta_{\infty}} $  and (c) $\omega= 0.81 \, \omega_{\Delta_{\infty}}$.
The gray lines in panels (b) and (c) represent the best fits of the numerical data with Eqs.~\eqref{eq:fit_1_mod} and~\eqref{eq:fit_1}, 
see discussion in the text.   }
\end{figure}

\begin{figure}[thp]
 \includegraphics[width=\columnwidth]{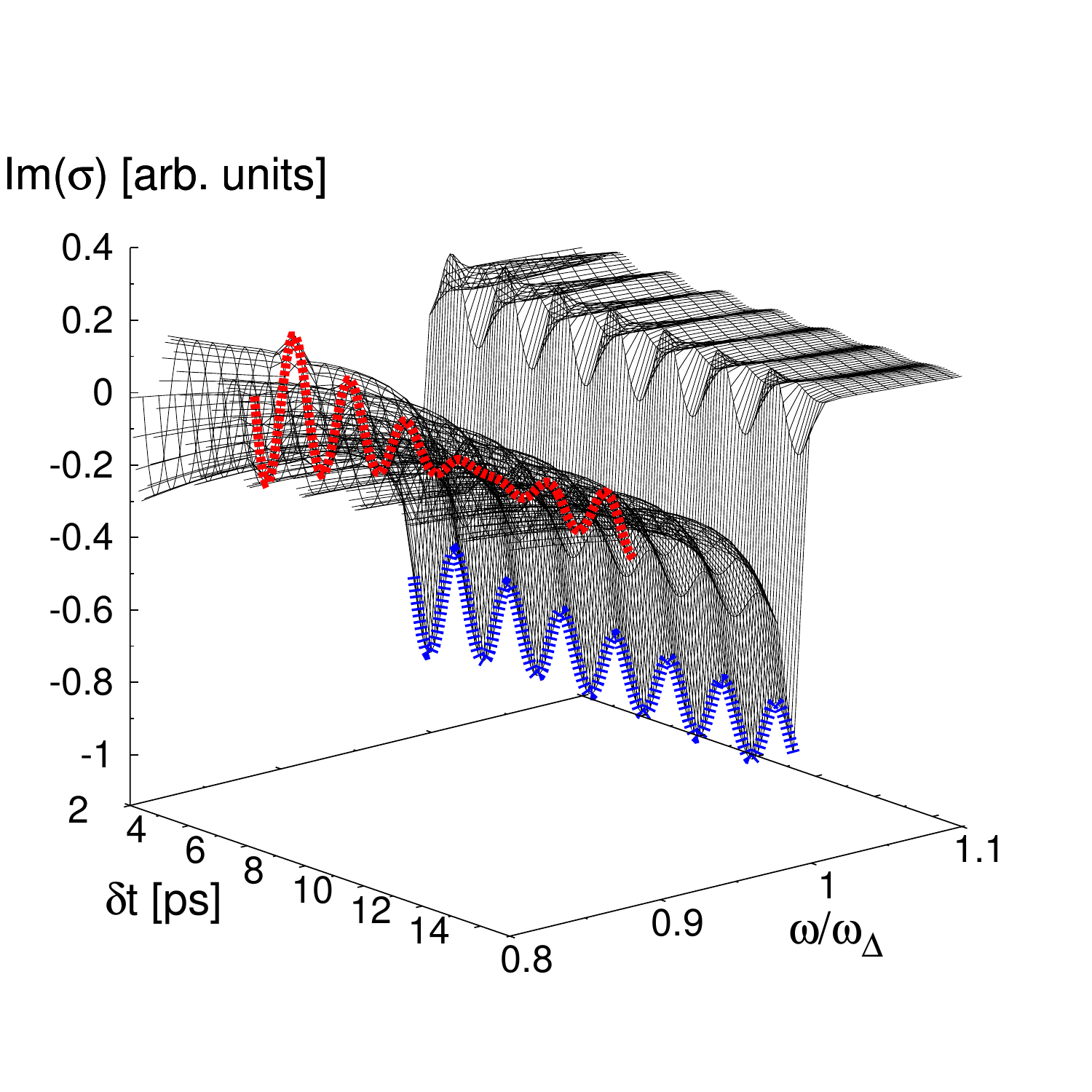}
  \caption{ \label{FigM12} \label{plot:sigma-im}
  (Color online) 
Imaginary part of the pump-probe signal, $\mathrm{Im} [ \sigma(\delta t,\omega) ]$, versus $\omega$ and $\delta t >0$ for the nonadiabatic regime [$\tau_{\textrm{p}}=0.5$ ps, $|\textbf{A}_{\textrm{p}}|=10\cdot10^{-8}$ Js/(Cm)] in the presence of an optical phonon mode with energy $\hbar \omega_{\textrm{ph}} = 2$~meV and
coupling strength $g_{\text{ph}}=0.1$~meV.  
 }
\end{figure}

\section{Imaginary part of\\
the pump-probe response}\label{sec:example_im}

For brevity, we have only presented  the results for the real part of the pump-probe response in the main text. Here, we show the results for the imaginary part, where we can observe the same signatures.

\subsection{Pump-probe response in the absence of phonons}

In Fig.~\ref{plot:sigma-im_gp_0}(a) we plot the imaginary part of the conductivity versus delay time $\delta t$ and frequency $\omega$ in the nonadiabatic regime with positive delay time. The corresponding real part is depicted in  Fig. \ref{FigM1}.  As discussed in the main text, we detect algebraically decaying oscillations as a function of the delay time.
For all frequencies except $\omega=\omega_{\Delta_{\infty}}$, the delay-time dependence can be described by Eq.~(\ref{eq:fit_1}). As an example, the  conductivity at  $\omega=0.81\omega_{\Delta_{\infty}}$ is shown in Fig.~\ref{plot:sigma-im_gp_0}(c). Here a perfect match of the fit and the data is shown.
The oscillation at $\omega=\omega_{\Delta_{\infty}}$, where we observe a sharp edge in $\sigma(\delta,\omega)$ as a function of $\omega$,  is governed by two frequencies, one is given by $\omega_{\Delta_{\infty}}$ and the other by $2\omega_{\Delta_{\infty}}$. This is due to the fact that the amplitude of the oscillation with frequency $\omega_{\Delta_{\infty}}$ is very small so that we can observe higher harmonics of this oscillation. To describe the delay-time dependence of the conductivity at this frequency, we have to modify Eq.~(\ref{eq:fit_1}) in the following way:
\begin{align}  \label{eq:fit_1_mod} 
&\textrm{Im} [ \sigma(\delta t , \omega_0) ] =\notag\\
&A+\frac{B_1\cos(\omega_{\Delta_{\infty} }\delta t+\Phi_1)+B_2\cos(2\omega_{\Delta_{\infty} }\delta t+\Phi_2)}{\sqrt{\delta t}}+C  \delta t.
\end{align}
The perfect match of this fit shows that this is a good approximation; see Fig.~\ref{plot:sigma-im_gp_0}(b).

\subsection{Pump-probe response in the presence of an optical phonon mode}

In Fig.~\ref{plot:sigma-im} the imaginary part of the conductivity versus delay time $\delta t$ and frequency $\omega$ is shown in the nonadiabatic regime with positive delay time near resonance. The corresponding real part is depicted in Fig. \ref{plot:sigma-rg1}. Also in the imaginary part we detect two edges in the frequency dependence of $\mathrm{Im} [ \sigma(\delta t,\omega) ]$, one at $\hbar\omega_{\Delta_{\infty}}=2.2491$~meV and one at the phonon energy $\hbar\omega_{\text{ph}}=2$~meV [blue and red traces in Fig.~\ref{plot:sigma-im}(a)]. The delay-time evolution of $\mathrm{Im} [ \sigma(\delta t,\omega) ]$ at $\omega_{\Delta_{\infty}}$ is again well described by Eq.~(\ref{eq:fit_1}) (not shown) . The delay-time evolution of $\mathrm{Im} [ \sigma(\delta t,\omega) ]$ at $\omega_{\Delta_{\infty}}$ is well described by Eq.~(\ref{fit_phonon_beating}) and shows the beating just as it occurs in the lattice displacement (not shown).
 In summary, the imaginary part of the conductivity shows the same signatures as its real part.


\bibliography{pumpprobe}

\begin{thebibliography}{39}
\bibitem[{foo({\natexlab{b}})}]{footnote1}
\bibinfo{note}{This saturation is due to the fact that according to Pauli's
  principle only one quasiparticle can be excited in each mode.}
\end{thebibliography}

\end{document}